\documentclass[prb,twocolumn,epsf,epsfig,amsmath,showpacs]{revtex4}
\usepackage[toc,page]{appendix}
\usepackage{hyperref}
\usepackage{graphicx,pdflscape}
\usepackage{color}
\usepackage{bm}
\usepackage{alltt,dsfont}
\usepackage{appendix}
\usepackage{amsmath,amssymb}
\usepackage{hyperref}
\hypersetup{
    colorlinks,
    citecolor=red,
    filecolor=cyan,
    linkcolor=blue,
    urlcolor=magenta
}

\usepackage{atveryend}
\makeatletter
\let\origcitation\citation
\AtEndDocument{\def\mycites{}%
  \def\citation#1{\g@addto@macro\mycites{#1^^J}\origcitation{#1}}}
\AtVeryEndDocument{\newwrite\citeout\immediate\openout\citeout=\jobname.cit
  \immediate\write\citeout{\mycites}\immediate\closeout\citeout}
\makeatother

\newcommand {\apgt} {\ {\raise-.5ex\hbox{$\buildrel>\over\sim$}}\ }
\newcommand {\aplt} {\ {\raise-.5ex\hbox{$\buildrel<\over\sim$}}\ }
\newcommand{\lessim}{\aplt}
\newcommand{\gssim}{\apgt}

\newcommand{\G}{{\cal{G}}}
\newcommand{\GH}{{\bf g}}

\newcommand{\tJ}{\ $t$-$J$ \ }
\newcommand{\beq}{\begin{eqnarray}}
\newcommand{\eeq}{\end{eqnarray}}
\newcommand{\barray}{\begin{eqnarray}}
\newcommand{\earray}{\end{eqnarray}}
\newcommand{\nn}{\nonumber}

\newcommand{\disp}[1]{Eq.~(\ref{#1})}

\newcommand{\refdisp}[1]{Ref.~(\onlinecite{#1})}
\newcommand{\figdisp}[1]{Fig.~(\ref{#1})}

\newcommand{\chem}{{\bm \mu}}

\begin{document}
\title{   Low energy physics of the  \tJ model in    $d=\infty$ using Extremely Correlated Fermi Liquid theory: Cutoff Second Order Equations  }
\author{ B Sriram Shastry }
\affiliation{Physics Department, University of California,  Santa Cruz, Ca 95064 }
\author{ Edward Perepelitsky }
\affiliation{Centre de Physique Th\'eorique, \'Ecole Polytechnique, CNRS, Universit\'e Paris-Saclay, 91128 Palaiseau, France}
\affiliation{Coll\`ege de France, 11 place Marcelin Berthelot, 75005 Paris, France}
\date{\today}
\begin{abstract}
{We present the results for   the  low energy properties of the infinite dimensional  \tJ model with $J=0$, using    $O(\lambda^2)$  equations  of the  extremely correlated Fermi liquid formalism. The  parameter  $\lambda \in [0,1]$ is analogous to the inverse spin  parameter $1/(2S)$ in  quantum magnets.
The present  analytical scheme allows us to approach the physically most interesting regime near the Mott insulating state $n\lessim 1$. It  overcomes the  limitation  to low densities $n \lessim .7$   of earlier calculations, by employing a variant of the skeleton graph expansion,    and a
 high frequency cutoff that is essential  for maintaining the known high-T entropy.   The resulting quasiparticle weight $Z$,   the  low $\omega,T $ self energy and  the   resistivity are reported. These   are  quite  close at all densities to the exact numerical results of the  $U=\infty$ Hubbard model, obtained using the  dynamical mean field theory. The present  calculation  offers the advantage of generalizing to finite $T$ rather easily, and allows  the visualization of the loss of  coherence of  Fermi liquid quasiparticles by raising  $T$. The present scheme  is     generalizable to   finite dimensions and a non vanishing $J$.
  }
 
 \end{abstract}
\maketitle

\section{Introduction}
{
 The  fundamental importance of the    \tJ model for understanding the physics of  correlated matter, including High Tc superconductors,   has been recognized for many years\cite{Anderson-tJ}. The \tJ model  is a prototype of extreme correlations,  incorporating the physics of (Gutzwiller) projection to the subspace of single occupancy.  The  added superexchange $J$  provides the mechanism for quantum antiferromagnetism at half filling,  and upon hole doping, for superconductivity via singlet pairing\cite{Anderson-tJ}. This viewpoint  has attracted much attention in the community.  It has led to many approximate methods of calculation  being applied to the \tJ model, in order to calculate experimentally  measured variables.   Despite intense effort in recent years,  schemes for  {\em controlled}  calculations  are rare, since the model has well known fundamental complexities that need to be overcome. 
 
 Motivated  by this challenge, we have recently formulated the  extremely correlated Fermi liquid (ECFL) theory\cite{ECFL-1,ECFL-11}, for tackling the \tJ and related $U\to \infty$ type models.  
 The ECFL  theory deals with the \tJ model by viewing it as  a non canonical Fermi problem, and proceeds via a non-linear representation of Gutzwiller projected Fermions in terms of canonical Fermions. It is  pedagogically useful to draw a parallel \cite{ECFL-11}  to the Dyson-Maleev representation of spins\cite{Dyson} used in  quantum magnets. In this representation\cite{Dyson}, the spins  are hard core Bosons,  and are non-linearly expressed in terms of  the canonical Bosons, namely the spin waves.
 The  ECFL methodology developed to date consists of successive approximations in the expansion   parameter $\lambda \in [0,1]$, playing a role  analogous  to the inverse spin parameter $1/(2 S)$  in quantum magnetism. { This analogy is developed in \refdisp{ECFL-11},
where parallels between the ECFL calculations  and earlier calculations of the partition function and Greens functions of the spin problem are drawn. It  is useful to note that the classical limit for spins $1/S \to 0$ corresponds to  the limit of free Fermion limit  $\lambda \to 0$.  Continuity in $\lambda$ leads to a protection of the Fermi surface volume for the interacting theory, i.e. the Luttinger-Ward volume theorem is obeyed. } Low order expansions can be performed analytically for most part, and therefore have all the usual  advantages of analytic approaches, such as explicit formulas for variables of interest and also flexibility for different situations. Several recent applications of the ECFL theory, mentioned below,  show  promise
in terms of reproducing the salient features of  exact numerical solutions of strong coupling models, wherever available\cite{ECFL-AIM,DMFT-ECFL}. The theory has also had success  in reconciling  extensive data on angle resolved photo emission (ARPES) line shapes\cite{ECFL-ARPES-1}, including subtle features such as the low energy kinks,  and  has made testable predictions on the asymmetry of line shapes\cite{ECFL-Asymmetry}.

In order to understand better the nature as well as limitations of a low order expansion in $\lambda$, we have tested the solution against two important strongly correlated problems where the numerical renormalization group and related ideas provide exact numerical results. In \refdisp{ECFL-AIM}, the asymmetric Anderson impurity problem, solved  by Wilsonian renormalization numerical  group  methods\cite{w,kww,Costi,Costi-Hewson} was used as one of the benchmarking models.  Secondly in \refdisp{DMFT-ECFL}, the $d\to \infty$ Hubbard model at large $U$, solved numerically by the Dynamical Mean Field Theory (DMFT)  method\cite{DMFT-1,DMFT-2,DMFT-3,DMFT-4,DMFT-5,DMFT-6,DMFT-7,DMFT-8,DMFT-9,DMFT-10,DMFT-11,DMFT-12,DMFT-13,DMFT-14,DMFT-15,DMFT-16,DMFT-17,DMFT-18,DMFT-19,DMFT-20,DMFT-21,DMFT-22,DMFT-23,DMFT-24,DMFT-25,DMFT-26,DMFT-27}, was used as the  benchmarking model.  These benchmarking studies show that the ECFL approach is overall  consistent with the exact solutions, with some caveats.  There are indeed differences in detailed structures  at higher energies\cite{comment1}. However the raw initial  results seem both useful and reliable for obtaining the low energy spectrum, and for a broad understanding of the occupied side of the spectral functions.  We further  found that the  calculation  are   {\em very close} to the exact solutions, {\em provided we scale the frequencies} by the respective quasiparticle weights $Z$  of the two theories.

The version of the ECFL presented in \refdisp{DMFT-ECFL} and the closely related \refdisp{ECFL-AIM}  is  therefore promising, but has the limitation  of being confined to low-density $n\lessim 0.7$\,. In the most  interesting density range $n\lessim 1$,   it falls short of being a ``stand-alone theory'', since the magnitude  of the calculated  $Z$ is too large. One  requires rescaling frequencies  to compensate for the incorrect magnitude of $Z$, and thereby  improve the agreement.
It is therefore  important to find ways to extend this analytical approach to cover the physically most interesting density regime $ .7 \leq n \leq1$.  A diagnostic objective  of this paper is to identify the cause for the inaccurate $Z$ in the earlier version, and to explore ways to overcome it. We have found it possible to do both.
This paper presents an alternative scheme that  can be pushed to high particle densities as well. We show here that   the resulting scheme gives satisfactory results for most of the interesting low $\omega,T$  variables of the model. 

Amongst the several variables of interest, the transport objects are  the most difficult ones to compute reliably. The difficulty lies in  their great sensitivity to the {\em lowest excitation energies}, and in the paucity of reliable tools to capture these. The limit of large dimensionality  is helpful here, since it has the great advantage of  killing the vertex corrections \cite{Khurana}. Thus  a knowledge of the one electron Greens function can give us {\em the  exact resistivity of a metal}, arising from inelastic mutual  collisions of electrons.
Despite the stated simplification, this calculation  remains technically challenging.
In important  recent work, this calculation has been performed   in \refdisp{Badmetal} and \refdisp{Kotliar}, for the large $U$ Hubbard model in infinite dimensions.
The authors  have produced     exact    resistivity results that are so  rare in condensed matter systems.    We can use them to  benchmark  our results for the resistivity at different densities and temperature. We   report   the results of this comparison in this paper. \figdisp{absolute-resistivity} shows one of the main  results of the calculation presented here, the details  leading to it are described below.
\begin{figure}[*htb]
\begin{center}
\includegraphics[width=.999\columnwidth]{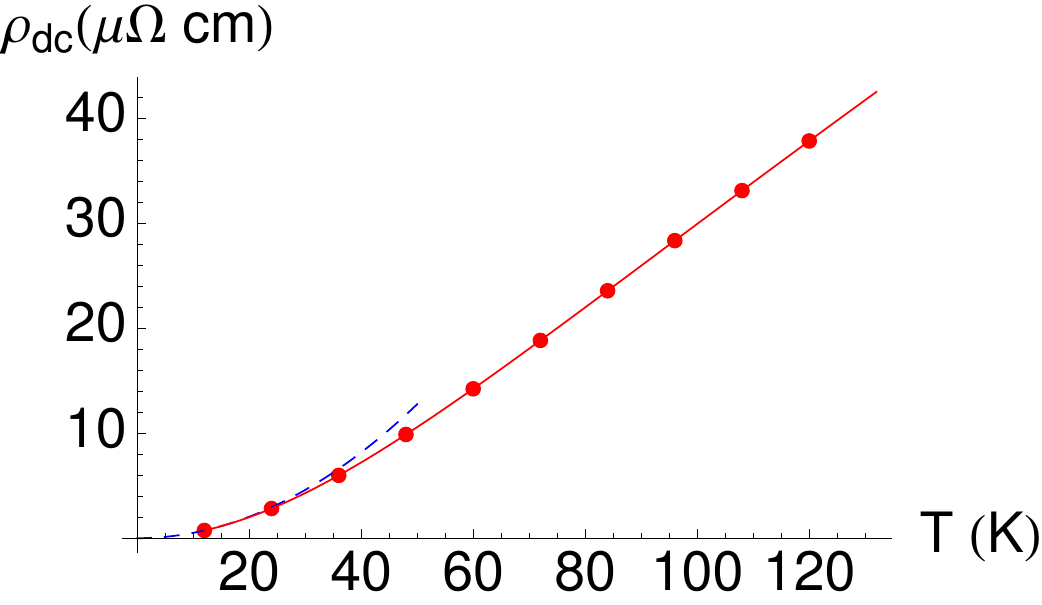}
\caption{$\rho_{dc}$ on absolute scale vs. $T$ in  Kelvin for particle density $n=.85$. We have used the estimates $D = 12000$K, and $\rho_0 =  258 \mu\Omega$ cm. The latter is obtained by using \cite{footnote-IRM} $\rho_0 \approx \frac{h a_0 d}{e^2}$, where we estimate $a_0 d\approx 10^{-8}$ cm. The Fermi-liquid behavior  with quadratic resistivity  in the blue dotted line,  breaks down above $T_{FL} \approx 30$K, and is followed by a regime of linear resistivity. } 
\label{absolute-resistivity}
\end{center}
\end{figure}

In Section \ref{I} we summarize the second order equations and introduce the various Greens functions and self-energies needed.  In Section \ref{II} we identify the conditions necessary for getting a satisfactory $Z$ near half filling. In Section \ref{III}  after summarizing the self consistency loop, we give a prescription for modifying the earlier equations and give the new set. This requires using a slightly different skeleton graph expansion, where certain objects are evaluated exactly using the number sum-rule. The ECFL theory has some intrinsic  freedom in choosing the details of the skeleton expansion, more so than in the  standard Feynman graph based canonical models.  That freedom can be usefully employed here. 
 We find that it is  also obligatory to introduce a high-energy cutoff, in order to recover  the known  high-T entropy of the model.  While the precise form of the cutoff is not uniquely given by theory, we found that several reasonable functional forms  gave comparable results at low energies and low T, provided that the  parameters were chosen to yield   the {\em high-T} entropy. 
 This cutoff  also  eliminates weak tails in the spectral functions that otherwise extend to large negative (i.e. occupied) energies.

In Section \ref{chempots}, we present results for the $T$ and $n$ variation of the  chemical potential and the quasiparticle weight $Z$. We also present  the $\omega$, $T$ and $n$ variation of the self-energy and spectral functions, where the quasiparticles, the asymmetry of the spectral functions and the  thermal destruction of the quasiparticles are highlighted. In Section \ref{resist} we present results for the resistivity at low and intermediate $T$ for various densities. In Section \ref{conclusions} we provide a  summary  and discuss the prospects for further work. 
}

\section{Summary of Second Order ECFL theory \label{I}}
Let us begin by recounting the exact formal expression for 
the Greens function  of the \tJ model. In the ECFL theory this object is given exactly as
 \beq \G(k,i \omega_n)=
  \GH(k,i \omega_n) \times \widetilde{\mu}(k, i\omega_n),  \label{eq7} \eeq 
 a product of the auxiliary Greens function $\GH$ and the   ``caparison'' function\cite{asyoulikeit}  given in terms of  a second self-energy $\Psi(k,i\omega_n)$ and   the particle  density  $n$ as $\widetilde{\mu}(k, i \omega_n) = \{1-n/2+ \Psi(k, i \omega_n)\} $.  The auxiliary Greens function $\GH(k,i\omega_n)$ given by
\beq 
\GH(k, i \omega_n)=\frac{1}{ i\omega_n+{ \chem} - \left\{ 1-n/2   \right\} \ \varepsilon_k-  \Phi(k, i \omega_n)
 }, \label{eq8} \eeq
where $\chem$ is the chemical potential and $\varepsilon_k$  the band energy.  In the infinite dimensional limit   it is demonstrated in \refdisp{Edward-Sriram-1} that an exact simplification occurs with these equations, whereby   the   momentum dependence is given by
\barray
\Psi(k, i \omega_n)&=& \Psi(i \omega_n),  \\
\Phi(k, i \omega_n)&=& \chi(i \omega_n) + \varepsilon_k \Psi(i \omega_n),
\earray
where both $\Psi$ and  $\chi$ are functions of only the  Fermionic Matsubara frequency
$\omega_n= (2 n+1)\pi \beta$,  {\em but not the momentum} $k$. These expressions  can be used in \disp{eq7} and upon
using the analytic continuation $i \omega_n \to \omega + i 0^+$, we may express the Greens function in the standard Dyson representation 
\beq
\G(k,\omega+ i 0^+)=\frac{1}{\omega+ i  0^+ + \chem - \varepsilon_k - \Sigma(\omega+ i 0^+)}\, ;
\eeq
where the Dyson self-energy  is now manifestly momentum independent, and given by
\beq
\Sigma(\omega+ i 0^+) =  \chem + \omega + \frac{\chi(\omega+ i 0^+) - \chem -\omega}{1-\frac{n}{2}+ \Psi(\omega+ i 0^+)}. 
\label{dyson}
\eeq
This result demonstrates the  momentum independence of the Dyson self-energy of the \tJ model in infinite dimensions. It  is consistent with the analogous   result  for the Hubbard model at any $U$\cite{DMFT-1,DMFT-2,DMFT-3}.

Within the ECFL theory we rely upon a systematic $\lambda$ expansion  to compute the two self-energies  $\Psi$ and $\chi$. This $\lambda$ expansion is  described in detail in  \cite{Edward-Sriram-2,ECFL-Monster, ECFL-1}, in brief the parameter   $\lambda$ lives in the range $\in \{0,1\}$, and  plays the role of the quantum parameter $1/(2 S)$ in the large spin expansions familiar in the theory of magnetism.  A skeleton diagram method can be devised for expanding the self-energies $\Psi$ and $\chi$ in a formal power series in $\lambda$, with terms that are  functionals of  $\GH$ and the band energies $\varepsilon_k$. This expansion uses the full $\GH$ (rather than  non-interactiong propagators $\GH_0$) as  fundamental units, or ``atoms'' for the expansion. The procedure is  in close analogy with the skeleton diagram methods used in many body theory. Having the self-energies to a given order in $\lambda$, one now reconstructs the Greens functions  self-consistently,  the scheme is to second order in the present case.

 The explicit equations to second order are found to be
\barray 
  \G(k,i \omega_n)&=&
  \GH(k,i \omega_n) \times \{a_G+ \lambda \Psi(k, i \omega_n)\} \label{eq7.2} \\
  \GH^{-1}(k, i \omega_n)& = & i\omega_n+{ \chem'} -  \lambda \; \chi(k , i \omega_n)\nn\\
  && - \{ a_G + \lambda
 \Psi(k, i \omega_n) \} \times (\varepsilon_k-\frac{u_0}{2}),\nn\\  
  \label{eq81} \earray
  with {
\beq
a_G= 1- \lambda \, \G(j,j^-) = 1-  \lambda \,  \sum_{k} \G(k, i \omega_n)\, e^{i \omega_n 0^+}, \label{eq819}
\eeq
where { $\chem'=\chem-\frac{u_0}{2}$}. In \disp{eq819} the middle (last) term is in space-time (wavevector-frequency) variables, denoted respectively  in  the  compact notation $j \equiv (\vec{R}_j, \tau_j)$,   $k \equiv (\vec{k}, i \omega_n)$, and denoting $j^- \equiv (\vec{R}_j, \tau_j + i\, 0^-)$.}
 The two self-energy  functions $\Psi$ and $\chi $  are expanded formally in $\lambda$ as
 $\Psi= \Psi_{[0]} + \lambda \Psi_{[1]}+ \ldots$ and $\chi=\chi_{[0]}+ \lambda \chi_{[1]}+ \ldots$.
 A systematic expansion in $\lambda$ is available to third order in \refdisp{Edward-Sriram-2}, from the low order results \cite{eqnreference}  we find $\Psi_{[0]}=0$, $\chi_{[0]}= - \sum_p \GH(p) (\varepsilon_p - \frac{u_0}{2})$ and  
\barray
 \Psi_{[1]}(k) & =& -   \sum_{pq}(\varepsilon_p+\varepsilon_q-u_0)\GH(p)\GH(q)\GH(p+q-k) , \nn \\
 && \label{psidef} \\
 \chi_{[1]}(k)&=& -  \sum_{pq}(\varepsilon_{p+q-k}-\frac{u_0}{2})(\varepsilon_p+\varepsilon_q-u_0) \nn \\
&&  \times \GH(p)\GH(q)\GH(p+q-k). \label{chidef}
\earray
In view of the explicit factors of $\lambda$ in Eqs.~(\ref{eq7.2},\ref{eq81}), 
this  leads to an $O(\lambda^2)$ approximation for $\G$;   the recipe further  requires that the parameter $\lambda$ is set to unity before computing. 
Here $u_0$ denotes  the second chemical potential.  It enters the theory as a Hubbard type term with a self-consistently determined coefficient $u_0$, as described in \refdisp{ECFL-Monster}. This chemical potential is essential in  order to satisfy the shift invariance of the \tJ model order by order in $\lambda$, namely  $t_{ij}\to t_{ij} + c \; \delta_{ij}$ with an arbitrary constant $c$. For instance we see in \disp{chidef} that a shift of the energies $\varepsilon_k \to c + \varepsilon_k $ is rendered immaterial due to the structure of the terms, the constant  $c$ can be absorbed into $u_0$. 
  The two chemical potentials $\chem$ and $ u_0$ are determined through the pair of sum rule on the auxiliary $\GH$  and the standard number sum rule on $\G$ 
\beq
\sum_{k} \GH(k)\, e^{i \omega_n 0^+} = \frac{n}{2}= \sum_{k} \G(k)\, e^{i \omega_n 0^+}\,  . \label{sumrule} \eeq
In  dealing with \disp{eq819} the composite nature of the $\G$  on view  in \disp{eq7},
offers a  choice  for  implementing the skeleton expansion. Such  a choice is absent in the more standard many body problems. On the one hand  we could  use the sumrule \disp{sumrule}  for $\G$ giving 
\beq
a_G^{(I)} \to 1- \lambda  \frac{n}{2},
 \label{exact-aG} 
\eeq
reducing to the exact answer $ a_G^{exact} = 1-  \frac{n}{2} $  as  $\lambda \to 1 $.

 Alternately  we could expand the $\G$ in powers of $\lambda$, a procedure we followed in \refdisp{ECFL-Hansen} and \refdisp{DMFT-ECFL}. We   expanded $\G$  out to first order in $\lambda$ from \disp{eq7.2}  since that   already gives the required  $O(\lambda^2)$ correction.  Thus we set  $\G = \GH (1-\lambda n/2) + O(\lambda^2)$, where the sum rule \disp{sumrule} was used for evaluating $\sum_{k,\omega_n} \GH(k)$. As a result we obtain the approximate result
\beq
a_G^{(II)} = 1- \lambda \frac{n}{2} + \lambda^2 \frac{n^2}{4} +O(\lambda^3). \label{approximate-aG} 
\eeq
Setting $\lambda\to1$ we thus get two alternate approximate skeleton versions of \disp{eq7.2} 
\beq
\G^{(I)}(k,i \omega_n) & =&  \GH^{(I)}(k,i \omega_n) \times \{1-  n/2 + \Psi(k, i \omega_n)\} \nn \\
&&\label{appx1} \\
\G^{(II)}(k,i \omega_n) & =&  \GH^{(II)}(k,i \omega_n) \nn \\
&& \times \{1-  n/2+  n^2/4 +  \Psi(k, i \omega_n)\} \label{appx2} ,
\label{eq9}
\eeq
where both expressions involve {\em the same} approximate $\Psi$ given in \disp{psidef},
and the auxiliary $\GH^{(.)}$   is also adjusted to have the appropriate expression for $a_G$ in \disp{eq81}. This dichotomous  situation arises due to the composite nature of the physical $\G$, whereas in standard many body problems the skeleton expansion is unique.

In \refdisp{DMFT-ECFL} as well as \refdisp{ECFL-Hansen} we employed \disp{appx2} to compute the electron self-energy and spectral functions. It was argued that this expression should be valid for low particle density $n \lessim 0.7$.
In \refdisp{DMFT-ECFL} the results were compared with the numerically exact DMFT results for the same model. It was found that  the self-energy is indeed close to the exact answer in the low-density limit. At the other end of {\em high-densities} $n \lessim 1$, it was   found  that the self-energy  is also very close to the exact result, provided we scale the frequencies by the quasiparticle weight  $Z$ of that theory. This remarkable observation shows that  in  ECFL theory, the Dyson  self-energy \disp{dyson} found by compounding two simpler expressions $\chi$ and $\psi$, has the correct functional form.  Moreover the  unusual and important  feature  of particle hole asymmetry, i.e. the presence of a strong $\omega^3$ term  in the $\Im m \, \Sigma$, comes about ``naturally'' within  the scheme. This feature has been argued to be generic for strongly correlated systems, as argued in \refdisp{ECFL-Asymmetry} and in the closely related  \refdisp{ECFL-AIM} for the Anderson impurity model.  The need for rescaling the frequency arises because the computed $Z^{(II)}$ using the approximate version \disp{appx2}, overestimates this variable as $n$ increases beyond the estimated limit of $n\sim 0.7$. We see in \refdisp{DMFT-ECFL} (Fig.16) that $Z^{(II)}$ does not even vanish as $n \to 1$, as one expects in a Mott insulator.

Within the spirit of \disp{appx2} one might expect that further approximations involving  higher order terms in $\lambda$ will enhance the range of validity in density. Such a program is essentially numerically intensive, since beyond second order one needs to use other techniques, such as  Monte Carlo generation and evaluation of diagrams \refdisp{QMC-1,QMC-2,Worm}. { We are currently performing these calculations, and have made formal progress towards this goal in \refdisp{Edward-Sriram-2}, by enumerating the non-trivial diagrammatic rules  in this model.  The diagrams that we encounter  include and go beyond Feynman diagrams, as necessitated by the lack  of Wick's theorem in 
 the non-canonical theory.}
 
On the other hand the analytical ease of the second order theory offers considerable advantage relative to other contemporary methods.  For low orders in $\lambda$ most calculations can be done by hand, and the remaining   computations  are modest in scope. Analytical methods 
also  have a  much greater flexibility, they can be applied in lower dimensions as well. Further the agreement with the other methods (DMFT \cite{DMFT-1,DMFT-2,DMFT-3,DMFT-4,DMFT-5,DMFT-6,DMFT-7,DMFT-8,DMFT-9,DMFT-10,DMFT-11,DMFT-12,DMFT-13,DMFT-14,DMFT-15,DMFT-16,DMFT-17,DMFT-18,DMFT-19,DMFT-20,DMFT-21,DMFT-22,DMFT-23,DMFT-24,DMFT-25,DMFT-26,DMFT-27}, numerical renormalization group \cite{ECFL-AIM}) and also experiments on ARPES for  the electron line shapes\cite{ECFL-ARPES-1}  is very good.   In view of these positive factors,  it  appears to  be  useful to  examine  if the   problem with the quasiparticle weight $Z^{(II)}$ at $n \lessim 1$  can be understood and corrected,  making other necessary approximations along the way.   This is indeed the purpose of this paper, we will see below that the  approximation \disp{appx1} provides us with the correct direction for such an approach.

\section{ The sum rules necessary for the vanishing of  Z near the Mott insulating state \label{II}}
\begin{figure}[b*thp]
\begin{center}
\includegraphics[width=0.9\columnwidth]{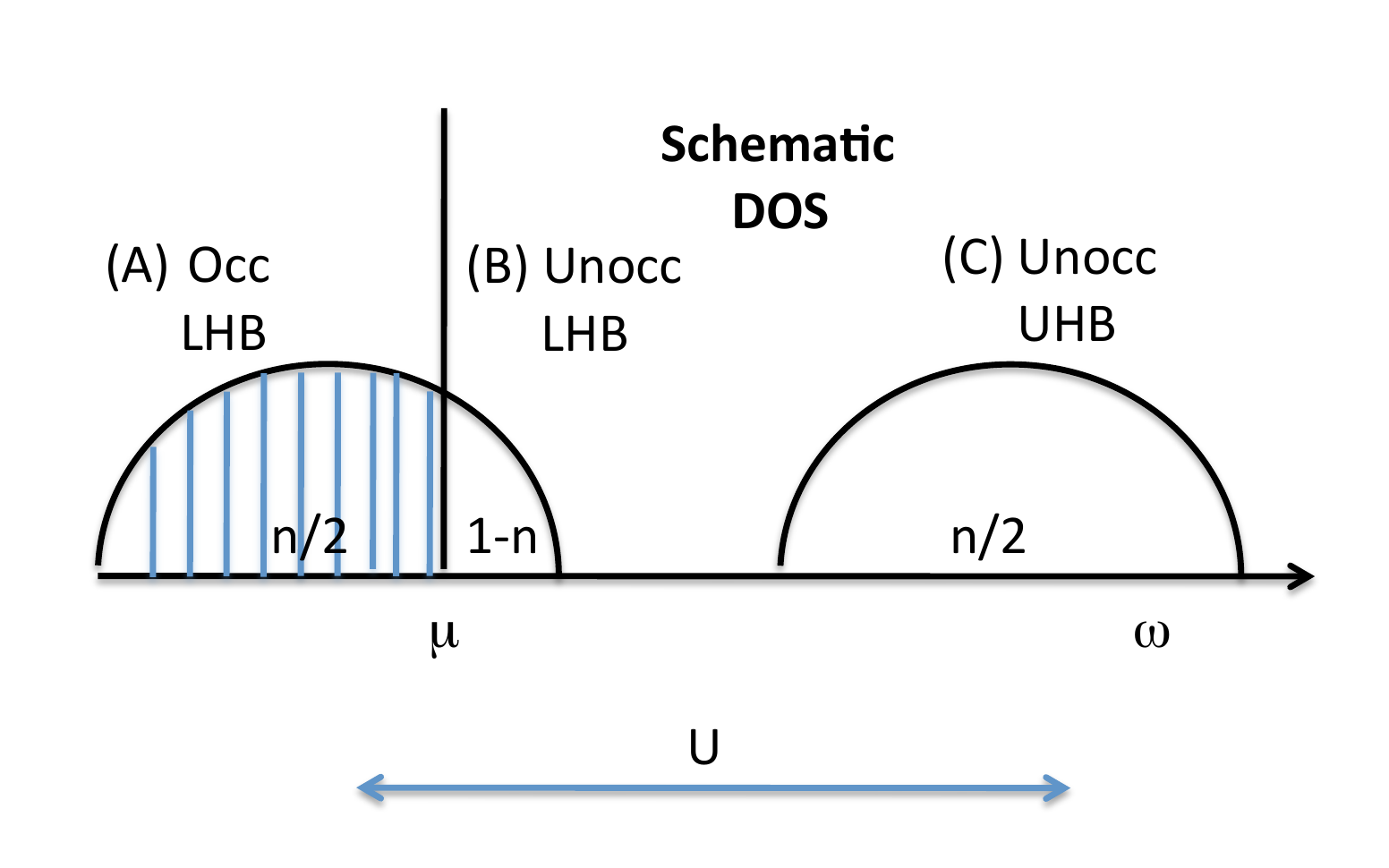}
\caption{\small \label{figspwt} A schematic depiction of the local spectral density-of-states $\rho_{G Local}(\omega)$ (popularly called $A_{Local}( \omega)$)   for the large $U$ Hubbard model, where the correlation split Hubbard bands are clearly separated. It  shows three regions (A) occupied electronic states (B) unoccupied lower Hubbard band states and (C) unoccupied upper Hubbard band states, with their respective weights as in \disp{srhubbard}. The \tJ model  sends  the region (C) off to infinity with weights given in \disp{sr3}. The area in region (B) is exactly $(1-n)$, and preserving this in an approximation  is key to obtaining the correct low energy scale.}  
\end{center}
\end{figure}

Let us first understand the factors that    make $Z$  vanish as we approach the Mott insulating limit. For this purpose it is useful to recall the local density-of-states of the Hubbard model  for the case of a sufficiently large $U$, (see \refdisp{phillips} for a useful discussion).  Here we expect the formation and clear separation of characteristic lower and upper  Hubbard bands - as indicated in the schematic \figdisp{figspwt}. 
Specializing to $T=0$ for simplicity, we note that  for the Hubbard model with $n<1$, the spectral weight for the local  $\rho_G(\omega)$ of the physical electron satisfies   the unitary sum rule $ \int d\omega \rho_G(\omega) =1$. We use a notation where  a sum over $\vec{k}$ is implied for unlabeled functions (without the $\vec{k}$ argument), e.g. 
$\rho_G(\omega) \equiv \sum_k \rho_G( \vec{k},\omega)$.
The local Greens function itself is given by
 \beq 
 G(\omega + i 0^+) =\int d \nu \frac{\rho_G(\nu)}{\omega - \nu + i 0^+},
 \eeq
 and so the $\omega \to \infty$ asymptotic behavior  is determined by this sum rule as $G(\omega) \to \frac{\int d\nu\rho_G(\nu)}{\omega } = \frac{1}{\omega}$. 
 This  can be partitioned into three  sum rules as depicted in \figdisp{figspwt},
 \begin{widetext}
\beq
 \int_{-\infty}^0 d\omega \rho_G(\omega) =n/2, \;\; \int_0^{\Omega_*} d\omega \rho_G(\omega) =1-n, \;\;  \int_{\Omega_*}^\infty d\omega \rho_G(\omega) =n/2, \label{srhubbard}
\eeq
\end{widetext}
where $\Omega_*$ is an energy scale  denoting  the upper end of the lower Hubbard band and hence is $\sim O(W)$- it is well defined provided $U \gg W$.  As stated these three integrals
add up to  1, ensuring that a full electron is captured.  On the other hand, the \tJ model spectral function $\rho_{\G}(\omega)$  satisfies
\beq
 \int_{-\infty}^0 d\omega \rho_{\G}(\omega) =n/2, \;\; \int_0^{\infty} d\omega \rho_{\G}(\omega) =1-n, \label{sr3}
 \eeq
 where the upper Hubbard band (and $\Omega_*$) is pushed off to $+\infty$, 
and thus the occupied and unoccupied portions add up to $1-n/2$. This can be visualized clearly with the help of \figdisp{figspwt}. This argument also determines the $\omega \to \infty$ asymptotic  form $\lim_{\omega \to \infty}\G(\omega) \to \frac{1-n/2}{\omega}$, and gives us a relation of importance to this study
\beq
&&\left(\lim_{\omega \to \infty}  \G(\omega)  \to \frac{1-n/2}{\omega} \right) \leftrightarrow \left( \int_0^{\infty} d\omega \rho_{\G}(\omega) =1-n \right)\nn \\
&&\label{conditionZ} 
\eeq
To see its relevance, we note that as $n\to 1$, the chemical potential  increases towards the top of the lower Hubbard band. This  implies that  the  unoccupied portion of the lower Hubbard band  shrinks to zero. 
Since  roughly half of the quasiparticle's weight \cite{half}  resides in this shrinking  energy domain of $O(1-n)$ times the band width, the quasiparticle residue $Z$ must vanish at least as fast as $O(1-n)$.

We may now refer back to \disp{appx2}; since from the definitions \disp{psidef} and \disp{chidef} we can see that   $ \lim_{\omega \to \infty}\left( \Psi(\omega),\chi(\omega) \right) \to 0$ and also 
$ \lim_{\omega \to \infty}\GH(\omega) \to \frac{1}{\omega}$, we combine these to obtain $$ \lim_{\omega \to \infty}\G^{(II)}(\omega) \to \frac{1-n/2+n^2/4}{\omega},$$
 whereby the unoccupied region  $  \int_0^{\infty} d\omega \rho_{\G^{(II)}}(\omega) =1-n+ n^2/4$, in conflict with the condition \disp{conditionZ} for a vanishing $Z$,  as $n \to 1$.

Having thus identified this weakness of the approximation, we also see by the same argument that \disp{appx1} would automatically give us  a vanishing $Z$, as $n\to1$; the factors are now appropriate for the condition \disp{conditionZ} to hold.

\section{  Cutoff second order ECFL theory \label{III}}

Motivated by the above discussion we  now implement  a skeleton graph expansion, where the basic atoms, or units,  are still $\GH$, but in static terms involving  $\G$, such as in \disp{eq819}, we use the exact particle number sumrule \disp{sumrule}. This leads us to study the equations in \disp{appx1}. 

\subsection{Full set of self-consistent equations}For convenience and future reference   we summarize the full set of equations to be solved self-consistently. These are similar to the ones used in
 \refdisp{ECFL-Hansen} and \refdisp{DMFT-ECFL} with all the necessary changes for the present case  made. The band  density-of-states is taken as the semicircular expression $D(\epsilon) = 2/ (\pi D) \sqrt{1- (\epsilon/D)^2}$, and thus $2D$ is the bare bandwidth.
  The complex frequency is denoted as $z= \omega+ i0^+$,   the local  Greens function and its energy moments are defined by 
\barray
\GH^{-1}(\epsilon,z) &=&  z + \chem' - (\epsilon- u_0/2) \left(1- \frac{n}{2} + \Psi_{[1]}(z)\right)\nn \\
&&  - {\chi}_{[1]}(z),  \label{s1} \\ 
\GH_{Loc,m}(z)&=& \int  d\epsilon \ D(\epsilon)  \GH(\epsilon, z) \times  \epsilon^m = \int  d\nu  \frac{\rho_{\GH L, m}(\nu) }{z-\nu}, \nn \\
&&\label{s2}
\earray
{
The chemical potential $\chem'$ absorbs all constants such as $\chi_{[0]}$,  leading  to
 \beq
 \chem = \chem' + \frac{u_0}{2} (1+ \frac{n}{2}) - \int d\omega\, f(\omega) \, \rho_{\GH L, 1}(\omega), \label{chemtrue}
 \eeq
 where $f(\omega)=1/(1+ \exp{ \beta \omega})$ is the Fermi function and we will need below $\bar{f}=1-f$.
The 
 \disp{s2} serves to introduce the spectral functions $\rho_{\GH L, m}(\nu)$, these are most often computed from the reversed relation 
 \beq 
 \rho_{\GH L, m}(\omega)= - \frac{1}{\pi}\,\Im m\, \GH_{Loc,m}(\omega+i 0^+ ). \label{spectralfunction}
 \eeq 
 The physical Greens function is found from $\G(\epsilon, z) = (1-n/2+ \Psi(z)) \times \GH(\epsilon, z)$, and the Dyson self-energy from $\Sigma(z)= z + \chem - \varepsilon - \G^{-1}(\epsilon, z) $.}  We define its local version $\G_{Loc}(\omega)$ and  its density through a band integration
 \barray
 \G_{Loc,m}(z)&=&  \int \, d\epsilon \, D(\epsilon) \times \epsilon^m \,  \G(\epsilon, z), \, \nn \\
 \rho_{\G L, m}(\omega) &=& -\frac{1}{\pi}\,  \Im m \, \G_{Loc,m}(\omega + i 0^+)\label{localphysical}.
 \earray
 The physical { momentum-integrated} spectral function $\rho_{\G L, 0}$ is an object of central interest. It is  also   needed for the number sum rule  below \disp{sumrulesagain}.  The computation of $\GH$ requires the two complex self-energies $\Psi,\chi$. These can in turn be found from expressions involving the fundamental convolution:
  \begin{widetext}
 \beq
\rho^{(\cal I)}_{a b c}( u) = \int_{u_1,u_2,u_3} {\small \delta(u+u_3-u_1-u_2) } \left\{ f(u_1)f(u_2)\bar{f}(u_3)+\bar{f}(u_1)\bar{f}(u_2)f(u_3) \right\} \times \rho_{\GH L, a}(u_1)\rho_{\GH L, b}(u_2)\rho_{\GH L, c}(u_3), \label{convolution}
\eeq
\end{widetext}
where the right hand side is  conveniently computed from the local densities $\rho_{\GH L, a}$, by using Fast Fourier transforms.
This density is required for  $(a,b,c)=0,1$, and determines the complex function
\beq
{\cal I}_{abc}(z) = {\cal P} \int d \nu \,  \frac{\rho^{(\cal I)}_{a b c}(\nu)}{z-\nu }.
\eeq
From this object  the two self-energies can be found as the combinations
\barray
\Psi_{[1]}(z)& = & 2 \, {\cal I}_{010}(z) - u_0 \, {\cal I}_{000}(z) \nn \\
\chi_{[1]}(z)& = & 2 \, {\cal I}_{011}(z) - u_0 \, \left({\cal I}_{010}(z) +{\cal I}_{001}(z) \right) + \frac{u_0^2}{2}\, {\cal I}_{000}(z). \nn \\
&& \label{selfenergies}
\earray
In summary we can compute $\GH$ in terms of $\chi,\Psi$ from \disp{s1}. Having done so we compute  $\chi,\Psi$ in terms of the $\GH$ from \disp{selfenergies},  thus defining the second part of the loop. The two chemical potentials $\chem$  and $u_0$ are found from 
\disp{chemtrue} and the two particle  number sum rules:
\barray
\int d \omega \, f(\omega) \, \rho_{\GH L, 0}(\omega) &=& \frac{n}{2}, \\
\int d \omega \, f(\omega) \, \rho_{\G L, 0}(\omega) &=& \frac{n}{2}, \label{sumrulesagain}
\earray
thereby all variables can be self-consistently calculated through a simple iterative scheme. The only inputs are  the density  of particles $n$ and the temperature $T$.

\subsection{Considerations of high-density $n \to 1$ at low T, and the entropy at  high T }
Before discussing the results, we note an important constraint that arises when we study the theory at high temperatures.  We need to make sure that the number of states after the Gutzwiller projection has the correct value, this requires that the chemical potential has the correct asymptotic value  at high $T$.   When $T\gg {t, J}$ the chemical potential grows linearly with $T$.   From  simple considerations of the atomic limit $t=0=J$,  one can calculate the partition function exactly, from this  one finds 
\beq
 \chem \sim k_B T \log\{n/(2(1-n))\}, \label{chemhighT} 
\eeq 
where $N_s$ and $n=N/N_s$ are the number of sites and the density respectively. This   linear growth with T with the correct coefficient also ensures that the entropy near the Mott limit  is correctly reproduced at high T.   Upon using  the Maxwell relation $(\partial S/\partial N)/_T = - (\partial \chem/ \partial T)_N$, and the intitial condition $S(n\to0)=0$, we find 
 \beq S\sim -k_B N_s \{ n \log{n/2} + (1-n) \log{(1-n)} \}, \label{entropy} 
 \eeq
 a well known result.   We must therefore also ensure that the approximation satisfies this condition \disp{chemhighT}, in order to obtain the correct entropy at high T. 
  
Upon solving the equations Eqs.(\ref{s1}-\ref{sumrulesagain}) at high densities $n \gssim 0.8$
 as $T\to 0$, or     high $T\gg D$ with moderated densitites $n \geq .7$,  we find that in each case the spectral function tends to flatten out on the occupied side, extending in range   to $\omega \ll - D$ with little weight in the tails. For the high T case  a second consequence is that the computed slope $d\chem/dT$ begins to depart from \disp{chemhighT}.
The flattening is consequence of the growth of $u_0$ which also increases linearly with $T$,
 becoming  larger than the bandwidth $2 D$, as seen in \figdisp{chempots1}. This  growth  enhances  the  coefficients in the self-energies \disp{selfenergies} and pushes one into a strong $u_0$ regime, unless we impose some cutoff.
  In the $T\to 0$  limit  the exact numerical results for  spectral functions from DMFT  \refdisp{DMFT-ECFL} do  confirm the expectation of a compact support for the spectral function, and hence the  observed growth is artificial. 

 \subsection{Cutoff scheme with a Tukey window }
\begin{figure}[bthp]
\begin{center}
\includegraphics[width=0.9\columnwidth]{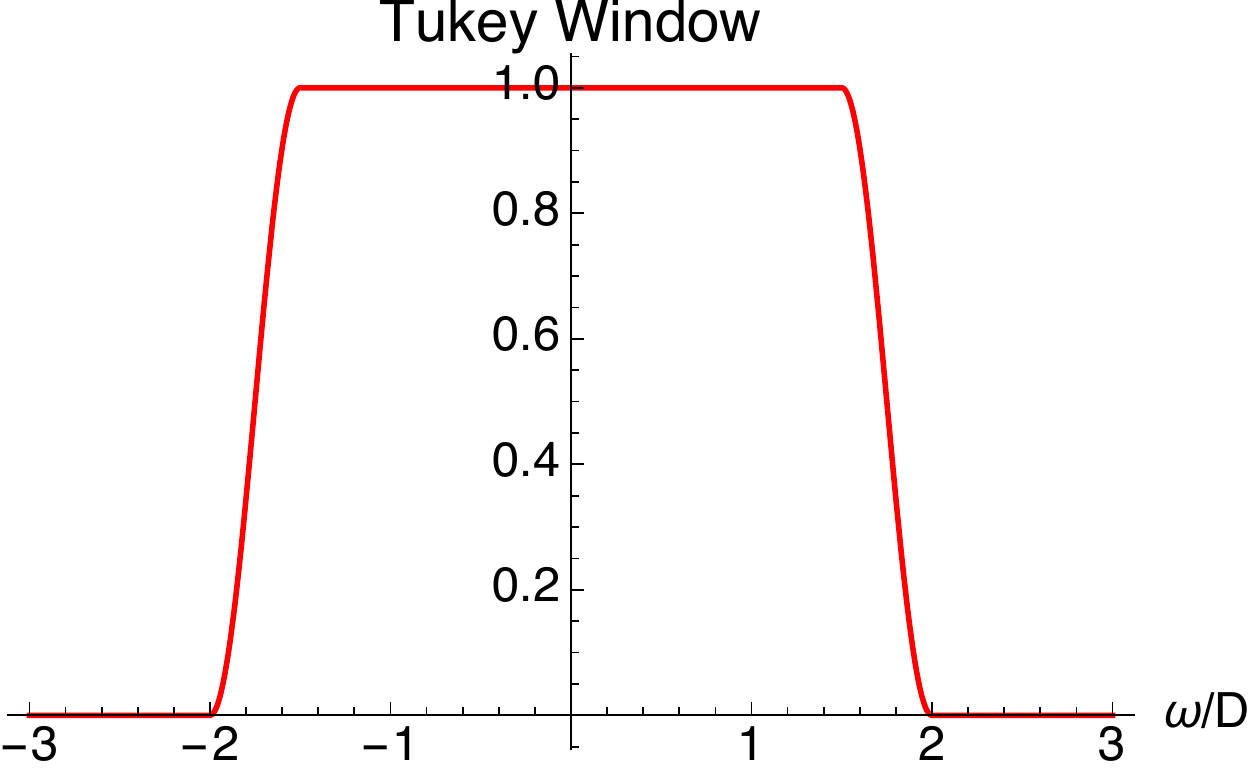}
\caption{ \label{Tukey} Multiplication through the  Tukey window ${\bf W}_T(\omega)$ (\disp{Wtukey}) is 
used for providing a cutoff   in our scheme \disp{auxrenorm}. It is applied only to the auxiliary  local Greens function $\rho_{\GH L, m}(\omega)$, while the physical spectral functions $\rho_{\G L}(\omega)$ are unconstrained, apart from an  overall window $|\omega| \leq 5D$ used for numerical purposes. 
In  this work,  the  upper cutoff used is
 $\Omega_c^{(+)}=2 D$, and the lower cutoff $\Omega_c^{(-)}=1.5 D$. }  
\end{center}
\end{figure}

We saw above    that  two physically  distinct  regimes involving  different types of physics, namely  the high T  regime  at any density   and the  high-density regime at low T share the common problem of growing tails of the spectral function.

In order to control this unphysical growth in both cases, we need to impose an appropriate high-energy cutoff.  Higher order terms in the $\lambda$ expansion are expected to eliminate this growth in a systematic way, without needing an extra prescription. { A detailed analysis of the cutoff issue within the $\lambda$ expansion is underway currently, and we expect to present the details in a forthcoming paper. } However  at the the level of the lowest order approximations, it seems that we do need to impose an extra cutoff- thereby introducing one more approximation.  A rough estimate of the cutoff can be made by observing that the self-energy calculated by using the bare $\GH_0$ (setting $\chi\to0$ and $\Psi\to 0$) in \disp{convolution} would give the spectral weights a width of maximum range $\pm 3D$; by setting $u_1=D,u_2=D,u_3=-D$ we satisfy one of the Fermi combinations with $u\sim 3 D$. By flipping signs we can reach $u=-3D$, thus a range of frequencies $-3D \leq \omega \leq 3 D$ is feasible.  The region near $|\omega| \sim  3 D$  would then be in the tails of the function. In a skeleton expansion on the other hand, with increasing interaction strength $u_0$, we have the possibility of a runaway growth,  since under first iteration,  the computed $\rho_{\GH L}$ can now extend to $\pm 3D$ as compared to the range $\pm D$ of the bare density, and so forth.  Hence one plausible strategy would be to limit the growth of the auxiliary spectral functions to a range { $\pm  c_0 $}, with $c_0 \sim$  2D,  with the physical spectral functions possibly extending somewhat beyond this.  Since two very different regimes, that of high-T and high-density are involved, we can test  the additional approximations self-consistently, and thereby avoid unduly biasing the results.

 It appears  reasonable to  choose the  high-energy  cutoff by requiring  that we obtain the  known   high T slope and therefore  the high T entropy \disp{entropy} at all densities. While it might be possible to obtain  the exact entropy by adjusting the cutoffs at each density separately, we content ourselves by finding a reasonable  global fit instead, i.e. one set of {\em density independent cutoffs} yielding  the  roughly correct entropy  at relevant densities.   The high T entropy is estimated  at   $T\lessim 1$. It should be noted that $T\sim 1$ is not always in the high $T$ limit, especially for the tricky region close to $n \sim 2/3$ where we know that $d\mu/dT$ vanishes at high T from \disp{chemhighT}, hence it is expedient to limit the high T region to $T\lessim1$.
  Having chosen such a cutoff, one can then explore the other physically interesting domain, and study  the spectral functions at low T  in the energy range $|\omega| \lessim D$.  This is a low energy scale compared to the cutoffs, but already a very high-energy scale, in comparison to   the physically interesting regimes $|\omega| \lessim \frac{D}{3}$ or even lower. We find below that the low $T$ spectra indeed are better behaved with the cutoff. The low energy results presented here are quite  insensitive to the details of the   choice for the cutoff, and hence one might be reasonably confident that the answers are not unduly biased by the choice made.
 
The  method   employed   for imposing  the high-energy cutoff was arrived at after some experimentation. We   multiply the local spectral function \disp{spectralfunction} by a Tukey window function used in data filtering:
 \beq
\hat{\rho}_{\GH L, m}(\omega) = \frac{1}{\cal V }\; {\rho}_{\GH L, m}(\omega) \times {\bf W}_{T}(\omega), \label{auxrenorm},
\eeq 
where the constant ${\cal V}$ is found from the normalization condition $\int \hat{\rho}_{\GH L,0}(\omega) \, d \omega=1$. Here the smooth Tukey  window  ${\bf W}_{T}(\omega)$ is unity over the physically interesting, i.e. feature rich  frequency domain $|\omega| \leq\Omega^{(-)}_c$, where it starts  falling  off smoothly, and  vanishing   beyond  the high  frequency cutoff $|\omega|=\Omega^{(+)}_c$.  It is defined as a piecewise function (see \figdisp{Tukey})
\begin{widetext}
\barray
{\bf W}_T(\omega)&=& 1, \, \mbox{for }\, \Omega_c^{(-)}\geq |\omega| \nn \\
&=& \frac{1}{2} \left( 1+ \sin \left\{ \pi/2 \frac{\Omega_c^{(+)}+\Omega_c^{(-)}- 2 |\omega|}{\Omega_c^{(+)}-\Omega_c^{(-)}}\right\}\right), \,\, \mbox{for }\,  \Omega_c^{(+)}\geq |\omega|\geq  \Omega_c^{(-)} \nn \\
&=& 0, \, \mbox{for }\,   |\omega| > \Omega_c^{(+)}. \label{Wtukey}
\earray
\end{widetext}
 This procedure involves {\em a single rescaling}: after computing the local spectral functions ${\rho}_{\GH L, m}$ (with m=0,1) from the self-energies as in \disp{spectralfunction},  we multiply with $W_T$ and rescale  as in \disp{auxrenorm}   before sending the result back into the self-energy calculation in \disp{selfenergies}. Note that the prescription \disp{auxrenorm} involves  the auxiliary local Greens function $\GH_{L,m}$ which is the basic building block in the theory.  The cutoff is imposed  {\em only} on $\rho_{\GH}$ in \disp{spectralfunction}, and the other spectral functions are then computed by the unchanged  Equations~(\ref{s1}-\ref{sumrulesagain}).

 We chose the parameters $\Omega_c^{(+)}=2 D$, and the lower cutoff $\Omega_c^{(-)}=1.5 D$ after some experimentation. This choice of the cutoffs is in accord with the discussion above where we concluded $c_0 \sim 2$D.
 With this  cutoff  and rescaled auxiliary Greens function,  the physical spectral function $\rho_{\G}$ 
 is computed as per the rules without any further assumptions. It typically
  {\em does} extend to about $4.5 D$ or $5 D$ on the occupied side, but not beyond this scale. For numerical purposes  we also use an upper cutoff for the physical  spectral function range  as $\sim 5 D$, this energy corresponds to $\Omega_*$ in \figdisp{figspwt}.

\section{Results for chemical potential, quasiparticle weight, self-energy and spectral functions  \label{chempots}}
\subsection{Chemical potential  and  quasiparticle weight $Z$. }
\begin{figure*}[*bht]
\begin{center}
\includegraphics[width=.9\columnwidth]{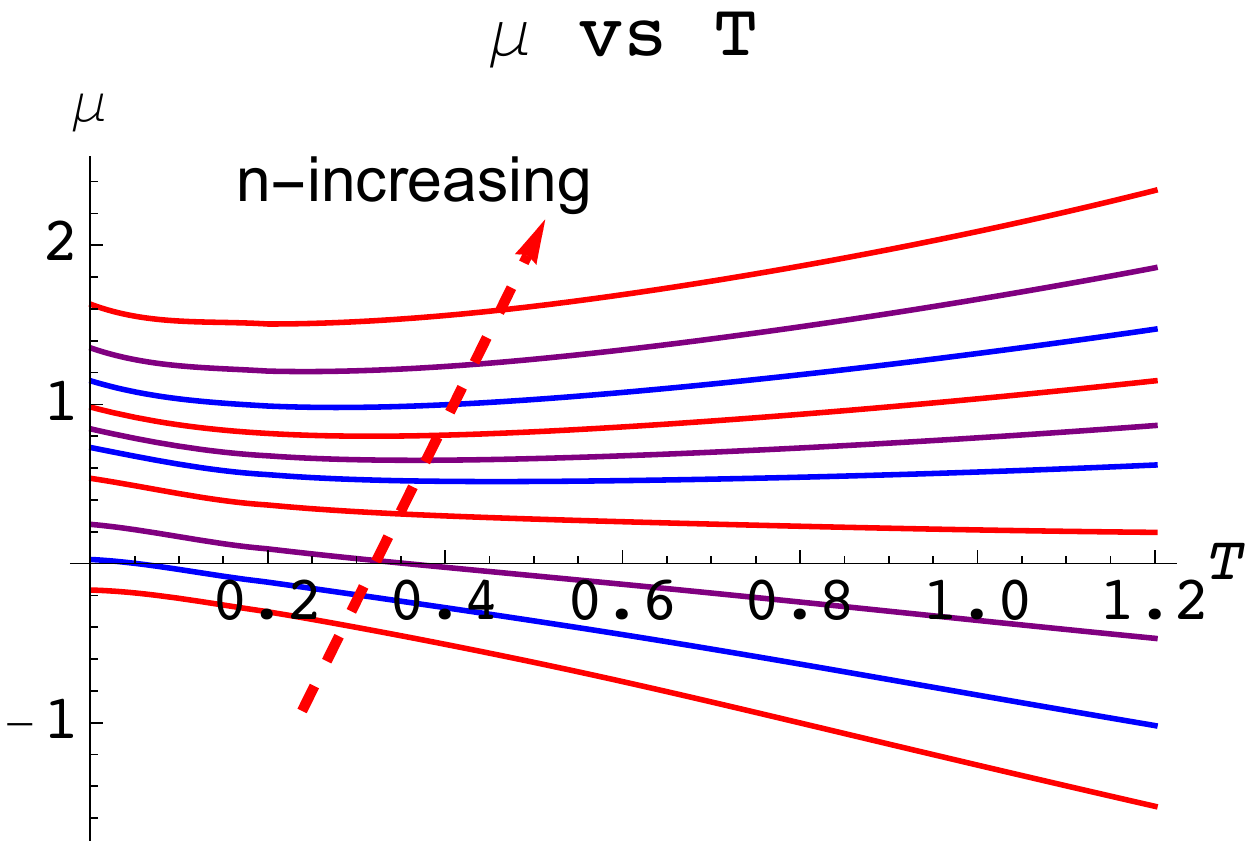}
\includegraphics[width=0.9\columnwidth]{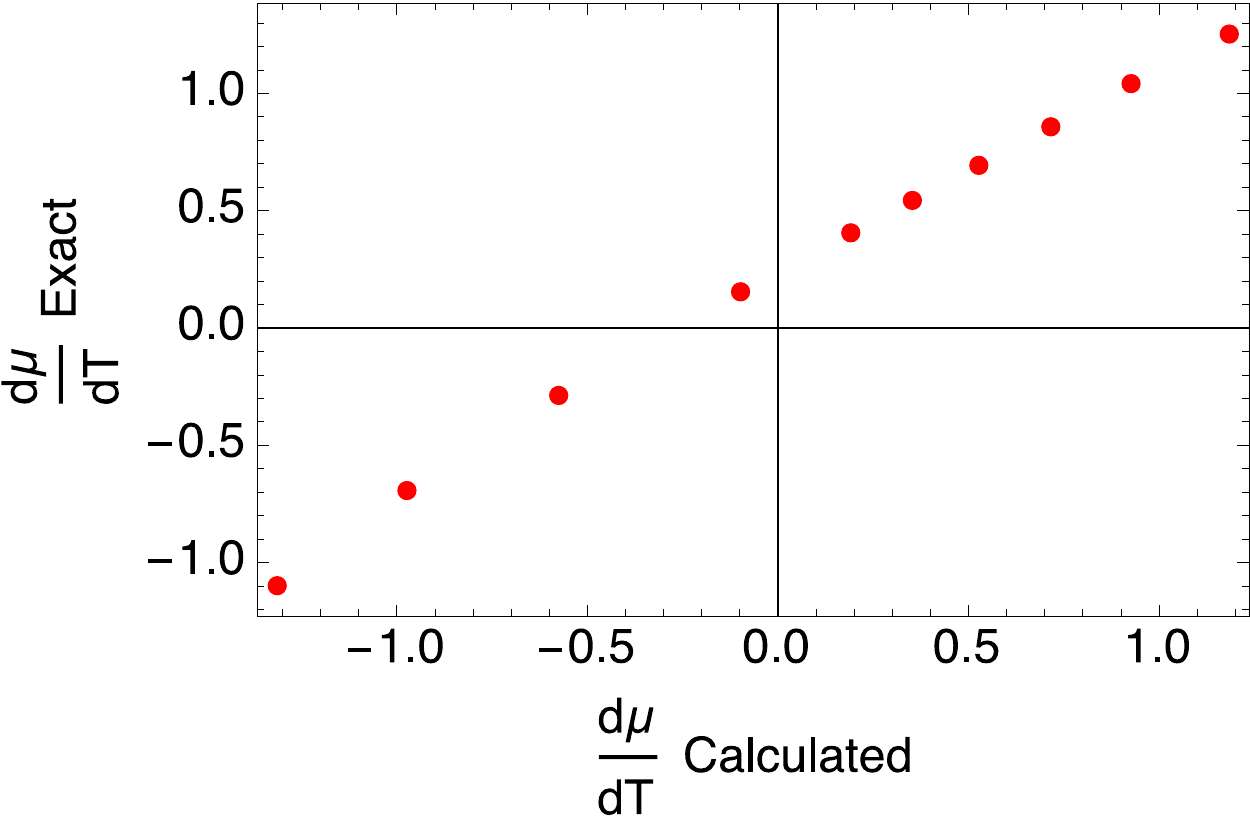}
\caption{{\bf Left}: The chemical potential at  particle  densities $n=.4,.5,.6,.7,.75,.775,.8,.825,.85,.875 $ increasing from bottom to top. We confine ourselves to the limited regime $T \leq 1.2$ since higher $T$ requires further adjustment of the cutoffs. 
 Note  that within this regime the  $\mu(T)$ curve  turns around  at a density around $n\sim .7$. For lower densities  $\mu$ decreases monotonically  with increasing $T$, whereas at higher densities we have a shallow minimum followed by a regime of rising $\mu$. This change of behavior is expected from \disp{chemhighT}, and  has  important physical consequence of changing the sign of the Kelvin thermopower for correlated matter \refdisp{Kelvin}.  {\bf Right}:  The slope $d\mu/dT$ is calculated  from the $\mu(T)$ curves  at $T=1$ , and 
 contrasted with the exact values from \disp{chemhighT}. The points are taken from the same set of particle densities $n$  as the figure on left, increasing from left to right.
 Since there is yet some curvature   in the figures at left when $T=1$,  our procedure provides only a rough estimate. We note that  these are  in fair correspondence, especially at low hole-density (see top right quadrant).
\label{muT}
} 
\end{center}
\end{figure*}

With the chosen cutoff we  examine the chemical potential  as a function of density and T in \figdisp{muT}.
We observe in the left panel of \figdisp{muT} that the chosen cutoff  provides a reasonable description of the $\mu$ vs. $T$ curves at different densities. These exhibit  an upturn  between $n=0.6$ and $0.7$ in the $T$ domain that is computationally reliable within this scheme.  The right panel of \figdisp{muT} shows that the slope $d\mu/dT$ is also in reasonable agreement with the exact answer for this slope,  apart from  some error near the difficult regime of $n\sim 2/3$.  Here we know from \disp{chemhighT} that the slope is zero at high enough T and this causes problems of convergence.

\begin{figure*}[bthp]
\begin{center}
\includegraphics[width=0.9\columnwidth]{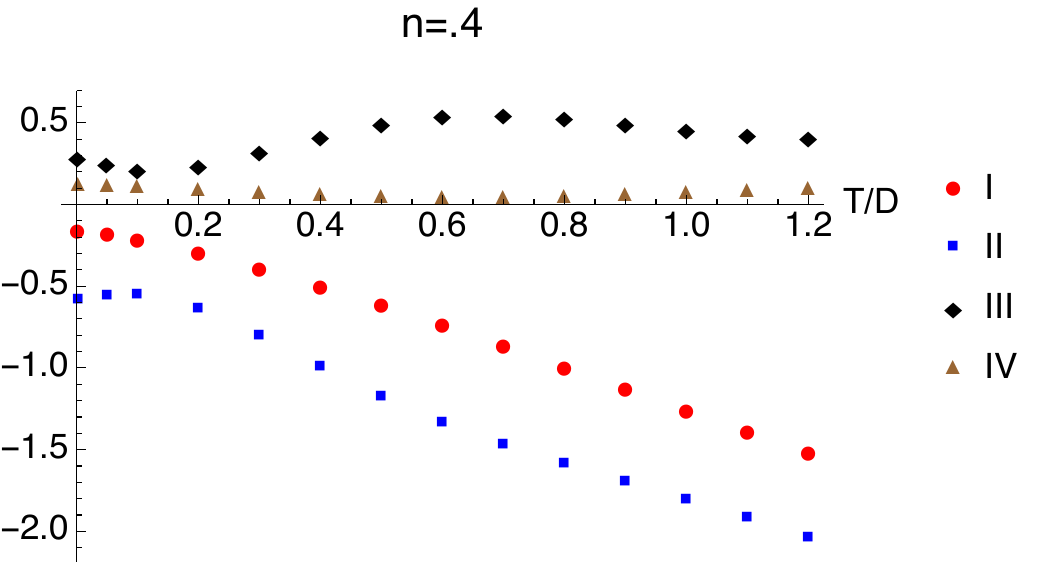}
\includegraphics[width=0.9\columnwidth]{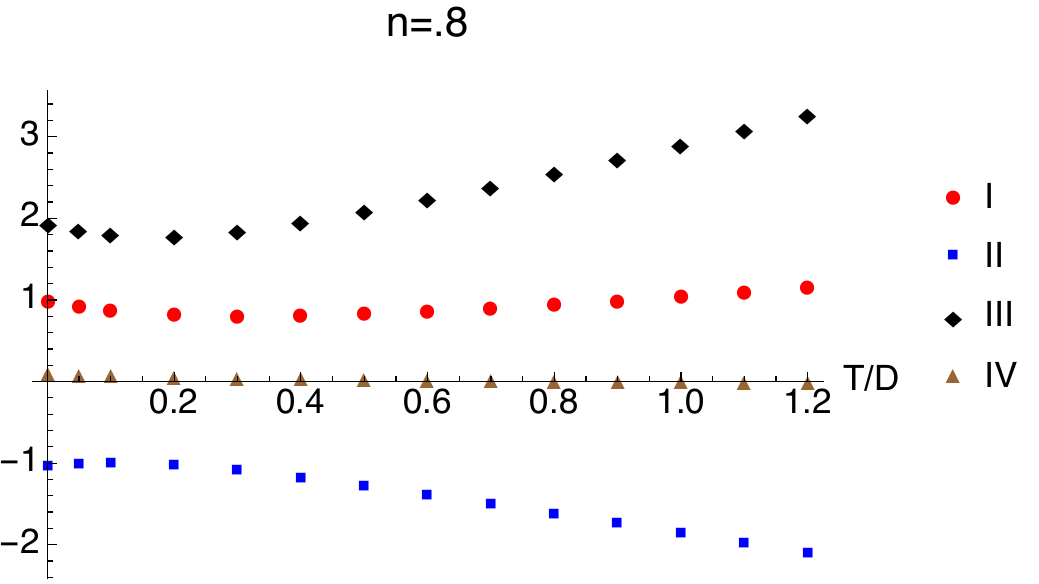}
\caption{\label{chempots1}The T dependence of the chemical potential $\chem$ and its three additive contributions from \disp{chemtrue} at two densities. The physical chemical potential $\chem$ (I-red), the auxiliary part: $\chem'$ (II-blue), the  $u_0$ contribution:  $(1/2+n/4) u_0$ (III-purple),  and the small part from the integral $-\int {f} \rho_{\GH L,1}$ (IV-magenta). The observed upturn in $\chem$ at high T for $n=.8$, reflecting the  physics of Mott holes near half filling, is predominantly due to the upturn of the second chemical potential $u_0$. Its growth, in turn, causes the numerical issues requiring the implementation of a cutoff in this work.
  }  
\end{center}
\end{figure*}

We  examine the various pieces adding up to the chemical potential in the right panel of \figdisp{chempots1}. 
These curves also show that the Mott-Hubbard physics of the upturn of $\mu(T)$ is enforced by the $u_0$ term, it is thus quite crucial within this formalism. We also note that  calculations without the cutoff lead to much larger values of $u_0$. 

 Overall  it  seems   that the results for $\mu$ are  quite reasonable in the hole rich region $n \geq .75$ (i.e. $\delta\leq .25$) with the  {\em global choice} made- i.e. without  requiring a fine tuning of the cutoffs with the density.  We therefore proceed to use this for computing the spectral functions, and other physically interesting variables, also evaluated in the {\em complementary low T region}. 
\begin{figure}[htb]
\begin{center}
\includegraphics[width=.9\columnwidth]{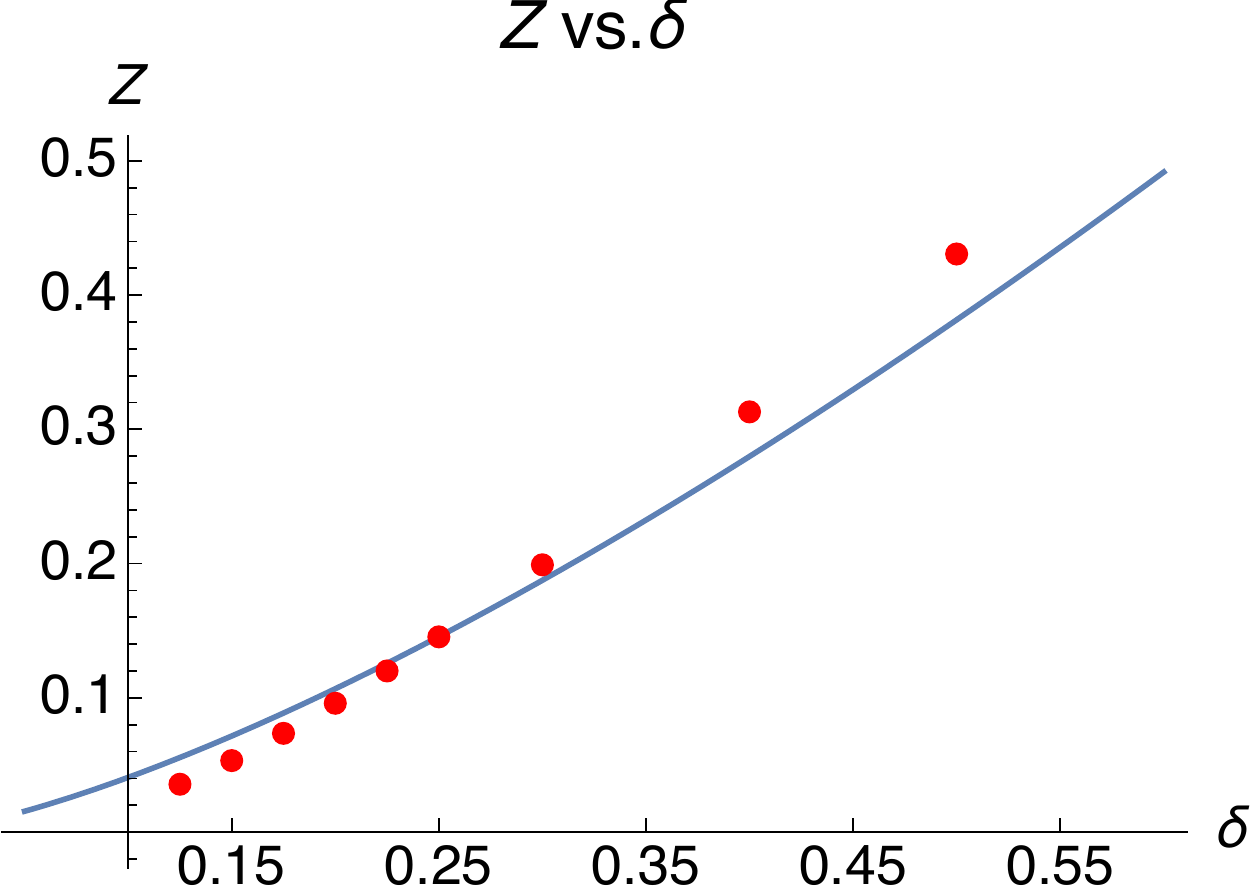}
\caption{  The computed quasiparticle weight  $Z $ (dots) versus the hole-density $\delta=1-n$, compared with the exact numerical results from DMFT (\refdisp{DMFT-ECFL} solid curve),  which   fits very well to the formula $Z\sim \delta^{1.39}$. We see that the present scheme accounts  well the suppression of $Z$ near $\delta \sim 0$,  even reproducing   non linear vanishing near the Mott limit seen in \refdisp{DMFT-ECFL}. This   nonlinear    feature  goes beyond the predictions of both slave Boson  mean field  and  Brinkman-Rice theory \cite{Brinkman-Rice}, and signifies an important correction to the mean field behavior. } 
\label{Zdelta}
\end{center}
\end{figure}

 Turning to the main objective of this work of  calculating the correct energy scale near the Mott limit, we display the computed $Z$ versus the hole-density $\delta=1-n$ in  the left panel of \figdisp{Zdelta}. It is interesting that the values obtained are significantly better than those reported in \refdisp{DMFT-ECFL}, we now find $Z$ vanishes as $\delta\to 0$. The solid line gives the numerically exactly determined $Z$ from DMFT, which is extremely well fit by $Z\sim \delta^{1.39}$.  This  latter behavior  is noteworthy  in that it  vanishes faster than linear in $\delta$. The ``mean field descriptions'' involving  slave auxiliary  particles as well as the Brinkman-Rice theory \refdisp{Brinkman-Rice} of the correlated metallic state give a linear $Z\propto \delta$. Therefore  this result indicates the need to account for fluctuations beyond the mean field description. It is interesting that the present calculation also gives a non linear behavior, with a slightly larger exponent than $1.39$. We plan to return to a closer analytical  study of this interesting result, obtained from the numerics of our solution.
 
 \subsection{self-energy and spectral functions at low T.}
\begin{figure*}[*htb]
\begin{center}
\includegraphics[width=.9\columnwidth]{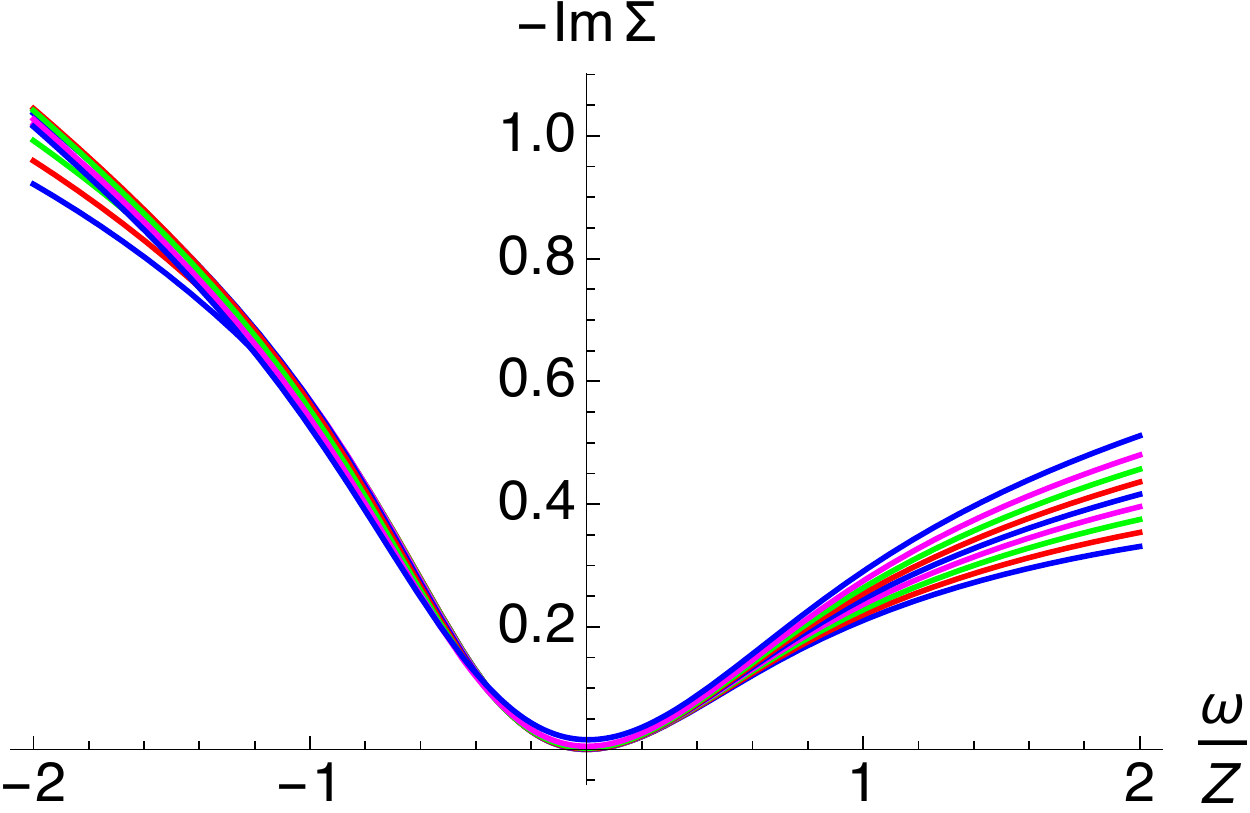}
\includegraphics[width=.9\columnwidth]{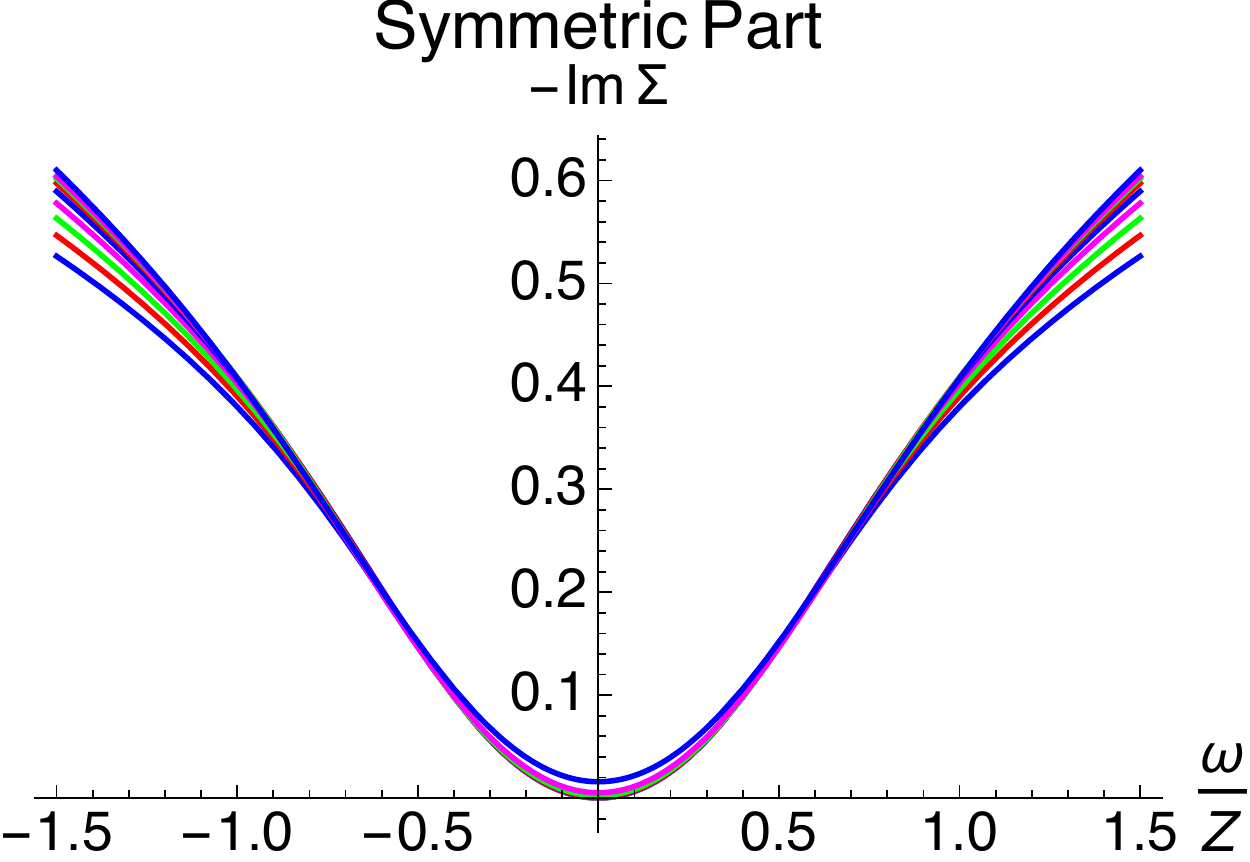}
\includegraphics[width=.9\columnwidth]{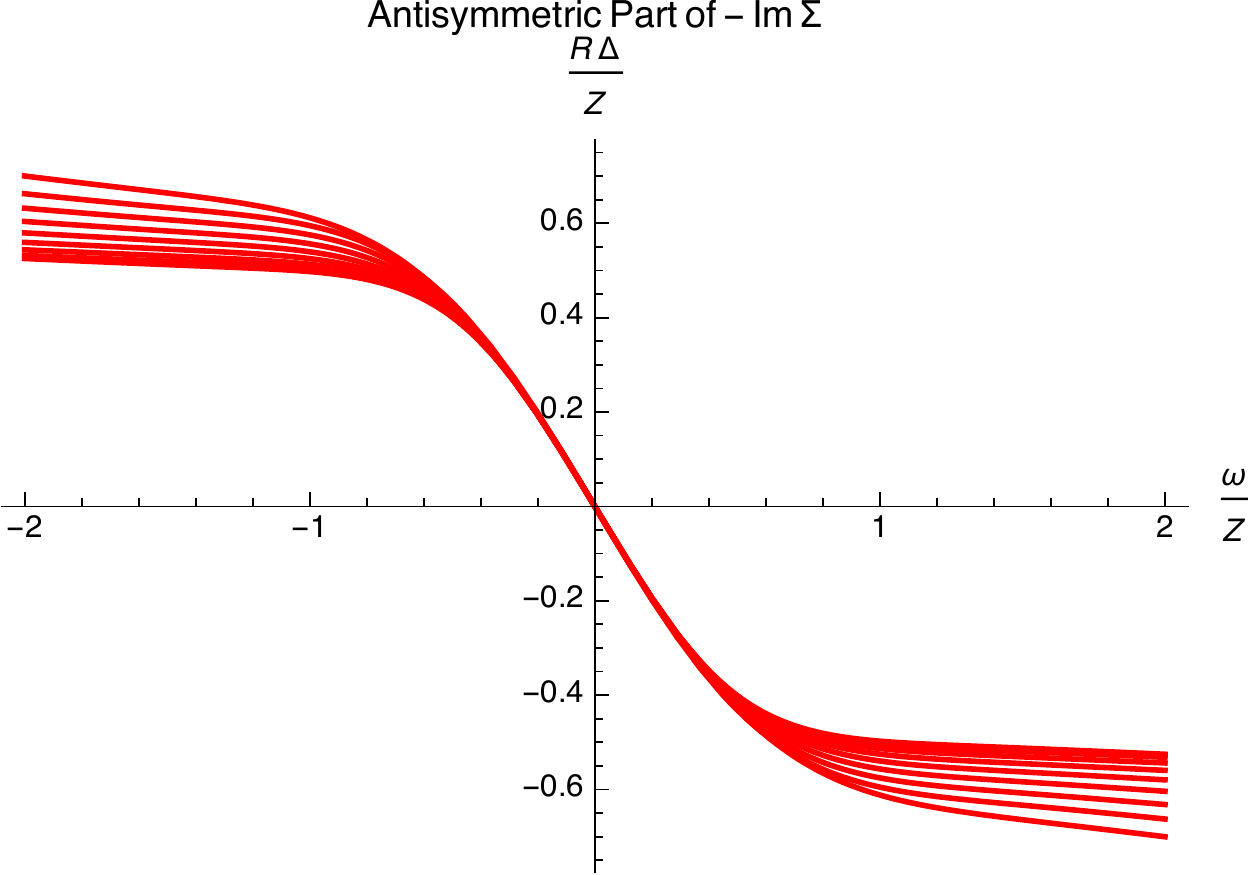}
\includegraphics[width=.9\columnwidth]{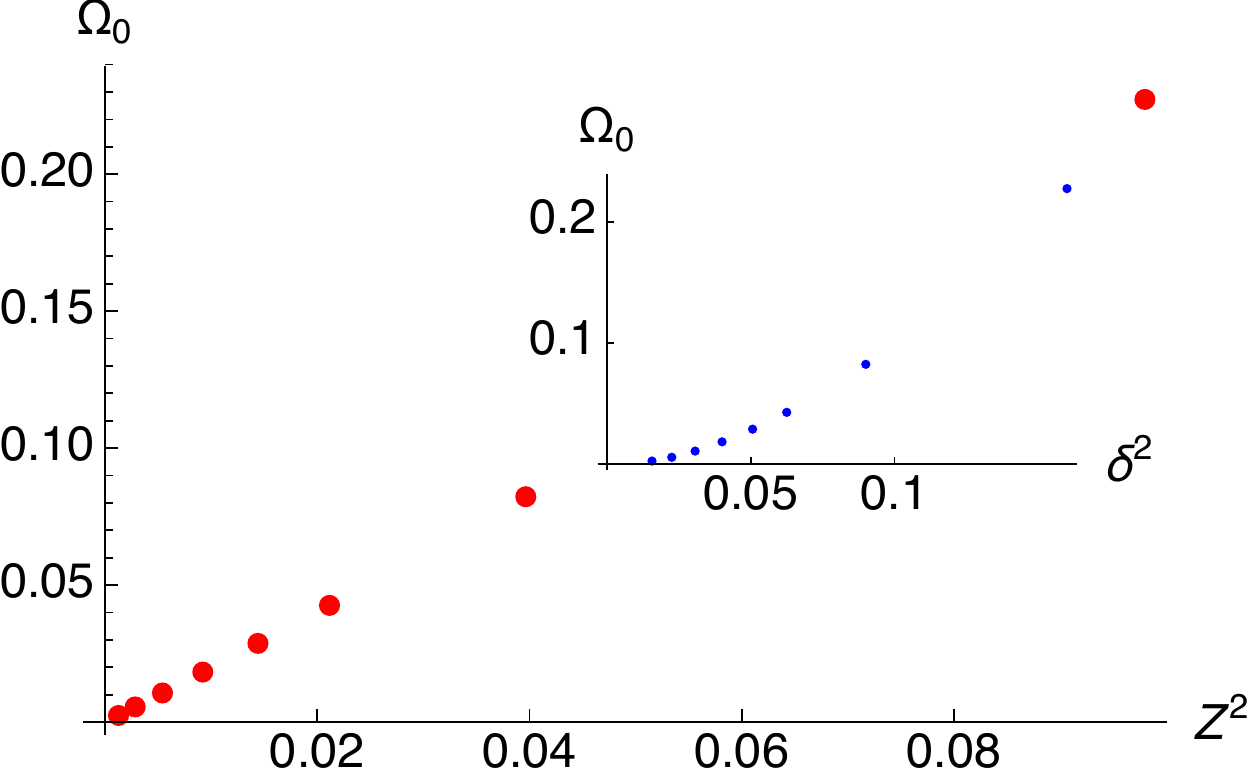}
\caption{ {\bf Top Left}:  $- \Im m\, \Sigma(\omega)$ versus $\omega/Z$ at several densities $n=0.7,.725,.75,.775,.8,.825,.85,.875,.9$ from bottom to top. We see that the frequency dependence scales well with $Z$, with better behavior on the occupied side $\omega \leq 0$. {\bf Top Right}: The symmetrized function $(- \Sigma''(\omega)- \Sigma''(- \omega))/2$ exhibits the   quadratic behaviour at $\omega\sim 0$ expected from a Fermi liquid.
  {\bf Bottom Left}: The antisymmetric part is defined as
 $R= ( \Sigma''(\omega)- \Sigma''(- \omega))/( \Sigma''(\omega)+ \Sigma''(- \omega))$, so that if we assume \disp{energyscales} then $R= -\omega / \Delta$.  We show the computed $R$  multiplied by $\Delta/Z$ at the above  densities versus $\omega/Z$, with $n=0.9$ at the top and $n=0.7$ at the bottom for $\omega \leq 0$. These 
  collapse to a  straight line with slope $-1$ in the range $|\omega|\leq Z$, provided we allow for an additional  mild   density dependence of the ratio $\Delta/Z$,   as in  \disp{Deltadep}.
   {\bf Bottom Right}:  The energy scale $\Omega_0$   \disp{energyscales} determining the magnitude of the $\Im m\, \Sigma$  at $T=0$ is shown versus  $Z^2$,  and  in the {\bf inset}   versus  the hole-density $\delta^2$. Here
 $\Omega_0$ is seen to scale better with $Z^2$   rather  than with $\delta^2$. } 
\label{DynamicSigma}
\end{center}
\end{figure*}

 We have also studied the quasiparticle decay rate at $T\sim0$,   
defined  for  $ |\omega| \leq Z D$ through a Fermi liquid form with the expected particle hole asymmetric correction\cite{ECFL-Asymmetry} 
  \beq - \Im m\, \Sigma(\omega) =  \frac{\omega^2}{\Omega_0} \times (1- \frac{\omega}{\Delta} )\label{energyscales}, \eeq
 whereby introducing two energies: $\Omega_0$, which determines  the  magnitude of $\Im m \, \Sigma$   and $\Delta$ the asymmetry scale. 
In \refdisp{DMFT-ECFL} and also in \refdisp{ECFL-AIM} it was pointed out that $\Omega_0$
 varies like $Z^2$ near the Mott insulating limit,  leading to a scaling of the Greens function frequency with $Z$ at low energies. In this work,
 the $\Omega_0$ is computed by  averaging  $\Im m\, \Sigma(\omega) $ in the domain $ |\omega| \leq Z D$.  In the bottom right panel of  \figdisp{DynamicSigma},  we show the variation of $\Omega_0$ versus $Z^2$ and in the inset with $\delta^2$.
 Since we have seen non linear corrections in $Z$ as seen in \figdisp{Zdelta}, these two plots  seem to support more closely   the scaling of $\Omega_0$ with $Z^2$, rather than $\delta^2$ at the lowest $\delta$. It seems possible to improve the agreement by choosing a density dependent  cutoff, however  the global cutoff already achieves  fair agreement.
 
In the top left panel of \figdisp{DynamicSigma} we plot  $-\Im m \Sigma$ versus $\omega/Z$ at different densities. As already noted in \cite{DMFT-ECFL}, these curves fall on top of each other
quite well. The curves also exhibit particle hole asymmetry as noted before  \cite{ECFL-1,ECFL-Asymmetry}. This is  exhibited by decomposing the $\Im m  \Sigma$ into symmetric and antisymmetric components in the the top right and bottom left panels. The antisymmetric part 
can be analyzed to read off the energy scale $\Delta$ in \disp{energyscales}. We find that $\Delta$ is proportional to $Z$ again,  but with a weak density dependent correction: 
\beq
\Delta(\delta)= Z(\delta) \times \left\{3.38 -15.6 \delta + 27.1 \delta^2 \right\} . \label{Deltadep}
\eeq
The region beyond the straight line is  captured on average,  by extending \disp{energyscales} to
 \beq - \Im m\, \Sigma(\omega) =  \frac{\omega^2}{\Omega_0} \times (1- \frac{\omega}{\Delta \sqrt{1+2\,\omega^2/Z^2} }).
 \eeq
This expression is potentially useful  for   phenomenological extensions of the theory.

\begin{figure*}[*bht]
\begin{center}
\includegraphics[width=.9\columnwidth]{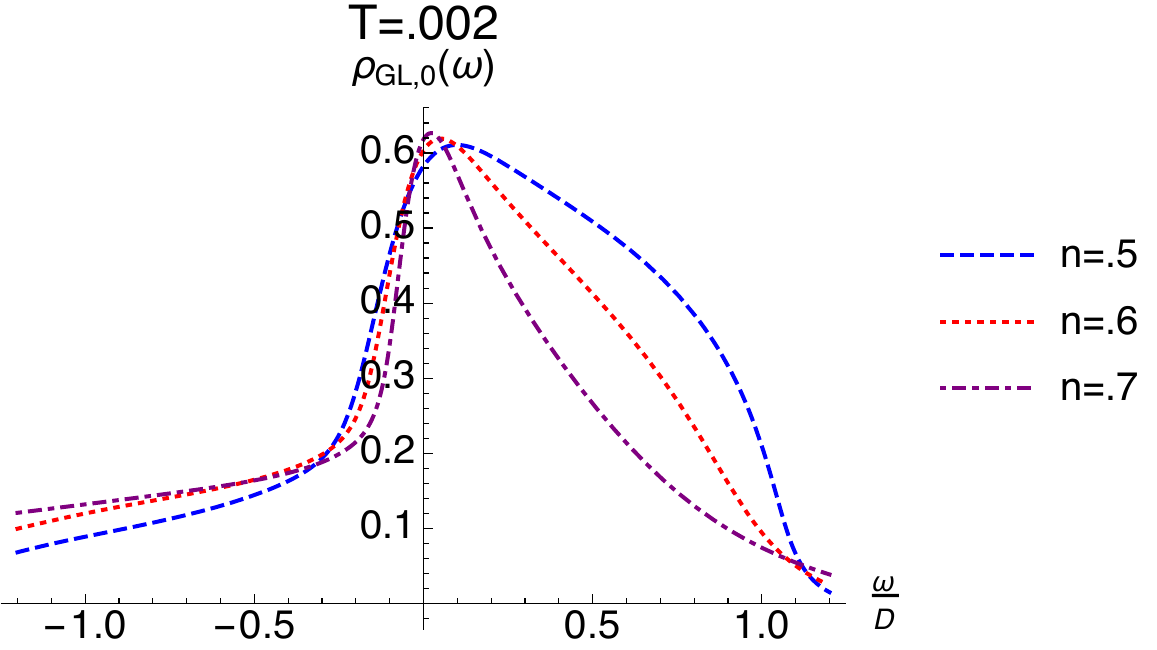}
\includegraphics[width=.9\columnwidth]{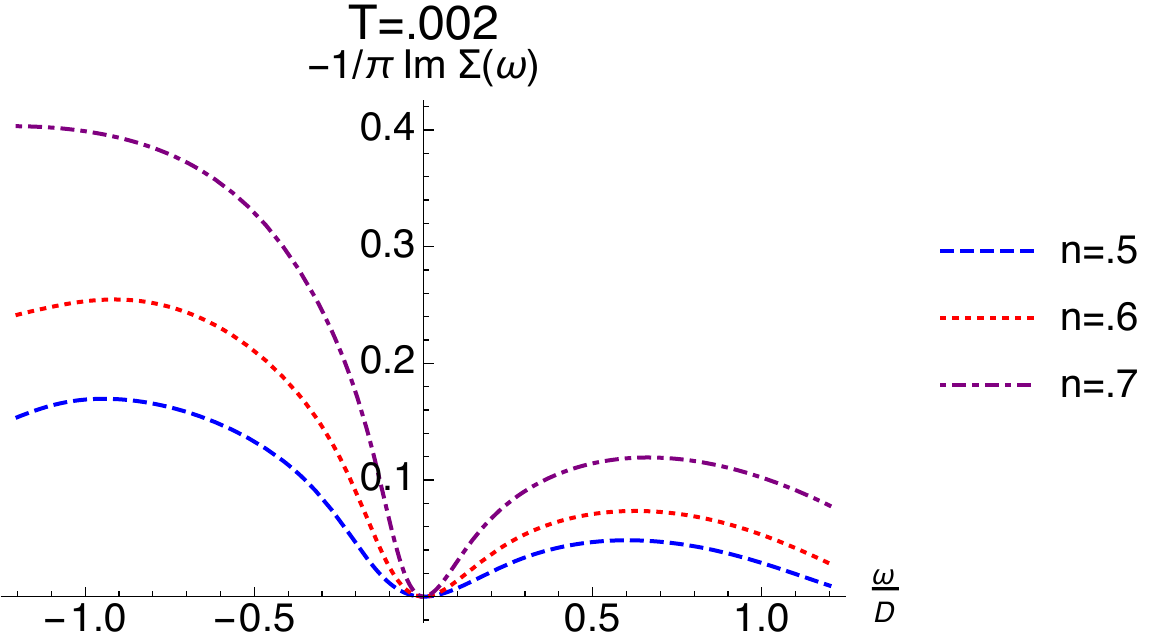}
\includegraphics[width=.9\columnwidth]{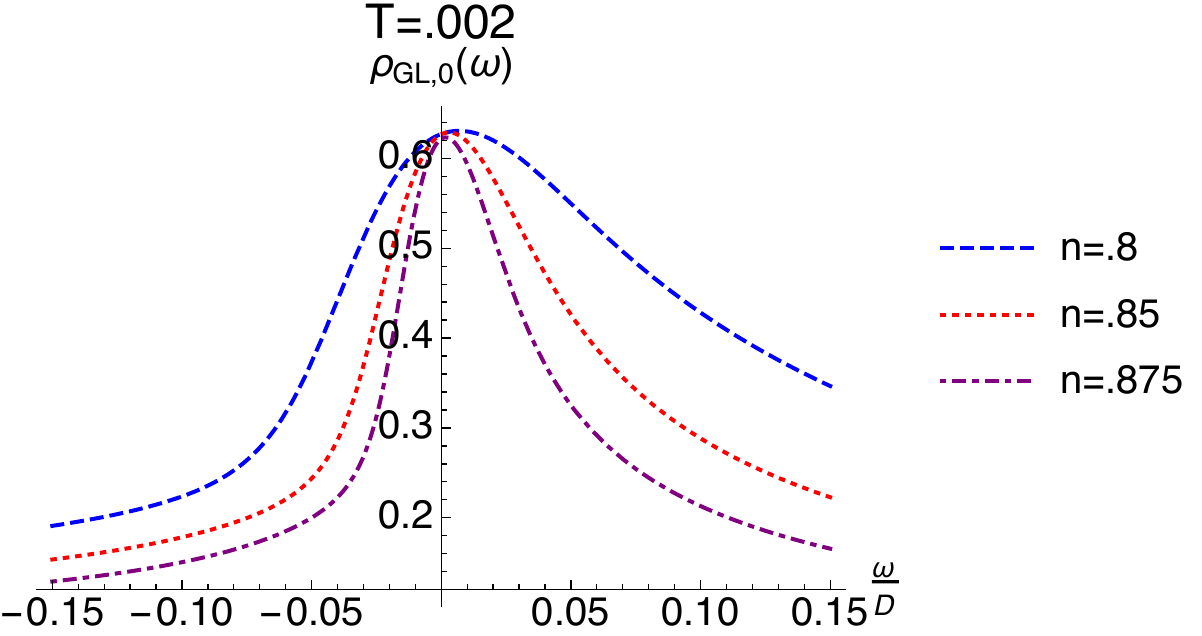}
\includegraphics[width=.9\columnwidth]{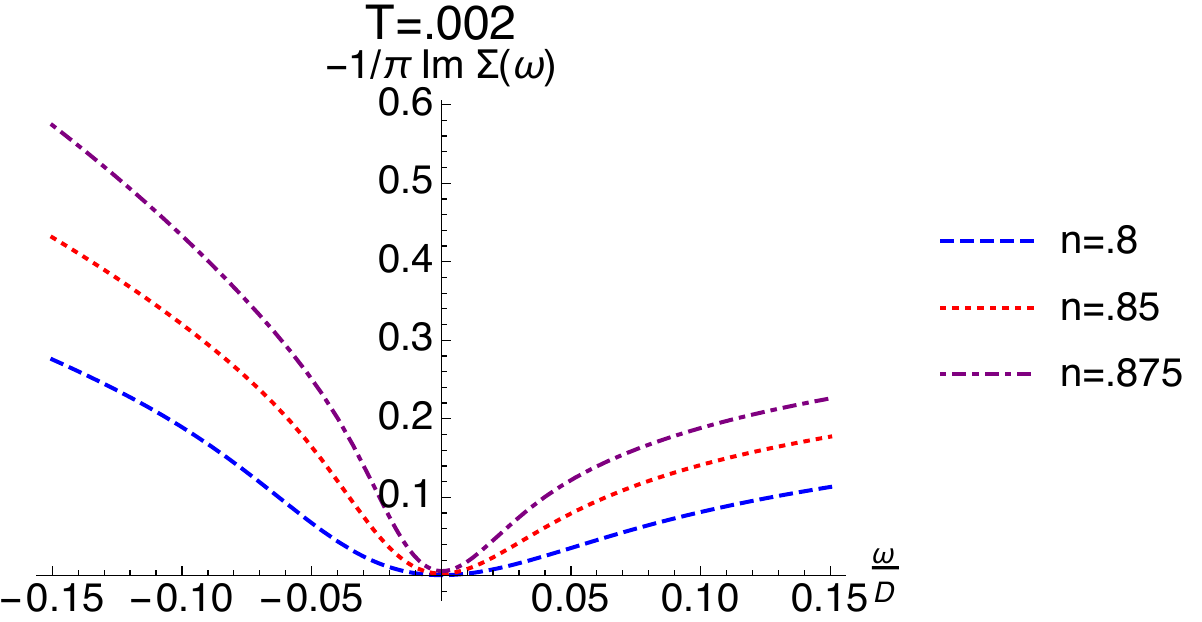}
\caption{ The  two figures  on the left display the physical  local spectral function $\rho_{\G L, 0}=  -\frac{1}{\pi}\,  \Im m \, \G_{Loc,0}(\omega + i 0^+)\ $ from \disp{localphysical},   and  the two figures on  right  show the Dyson self-energy  $- \frac{1}{\pi} \Im m \, \Sigma(\omega)$, plotted against the frequency $\omega/D$. The figures are at low $T$ for the six indicated values of the density, and  display   a region that is somewhat  greater than the one, where it is  expected to be reliable $|\omega| \lessim { Z D}$.  One sees a correlation between the quasiparticle weight  $Z$ (\figdisp{Zdelta}) and the scale of variation of the decay rate. { Densities $n> .875$ have larger errors in $Z$ compared to the exact DMFT results (see  \figdisp{Zdelta}), and therefore not shown. However  it is easy to picture them at low $\omega$, using the observation that scaling $\omega$ with $Z$   collapses  $\Sigma''$. }
} 
\label{LowTspectra}
\end{center}
\end{figure*}

In \figdisp{LowTspectra} and \figdisp{HighTspectra}, we display the  raw unscaled spectral functions and the imaginary part of the  self-energy  for  various physical parameters. In \figdisp{LowTspectra}  the low $T$ spectra are shown at different densities.  Note that the significant range of $\omega$ where the spectral  functions and self-energy vary, shrinks rapidly with increasing $n$- this is indirectly a reflection of variation of the $Z$ with density in \figdisp{Zdelta},  since the scale of variation of $\Sigma$ is set by $Z$. We also note that the spectral asymmetry in $\Im m \Sigma$ is very clearly visible here.

\subsection{Temperature variation of the self-energy and spectral functions.}

\begin{figure*}[*htb]
\begin{center}
\includegraphics[width=.95\columnwidth]{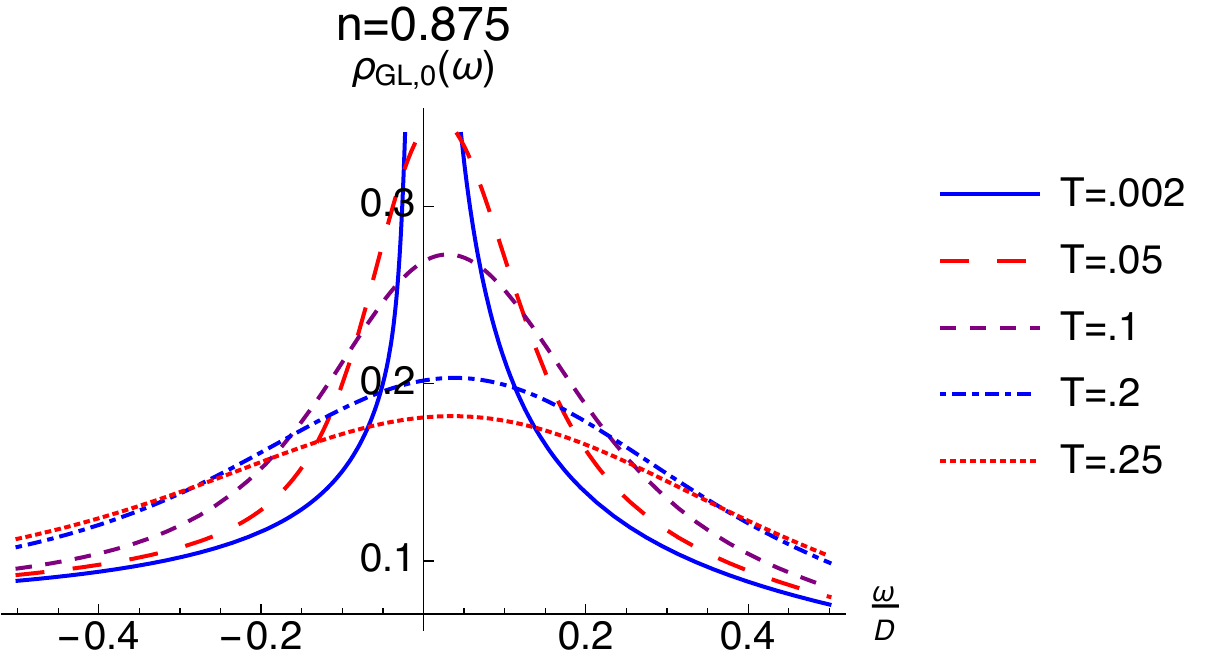}
\includegraphics[width=.95\columnwidth]{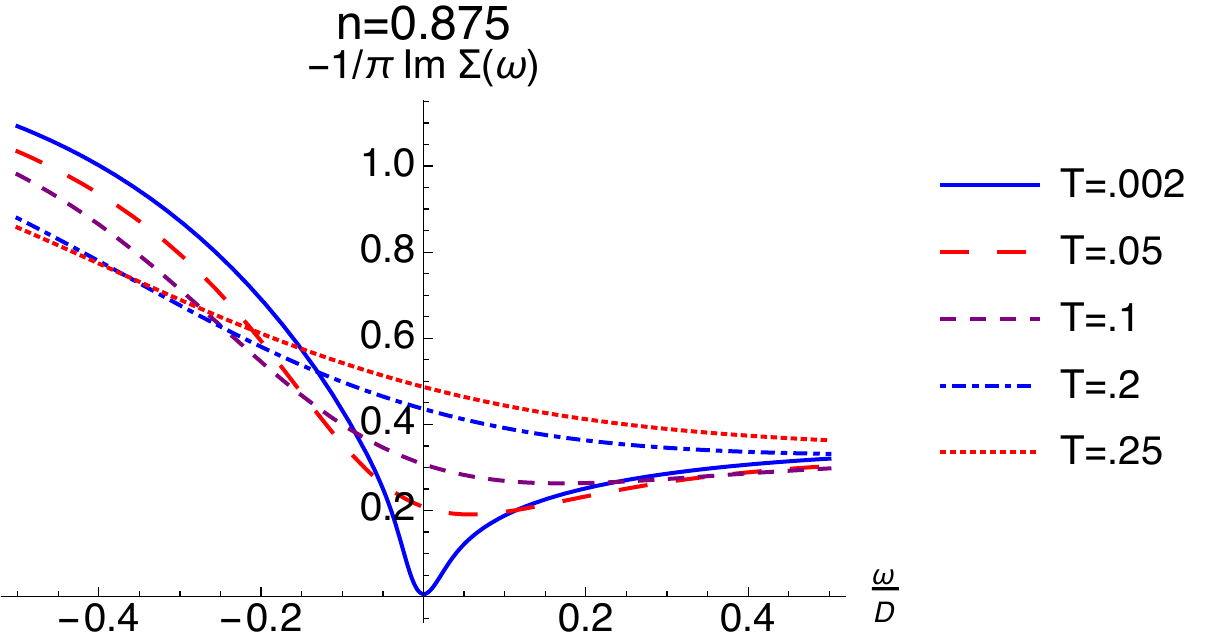}
\includegraphics[width=.95\columnwidth]{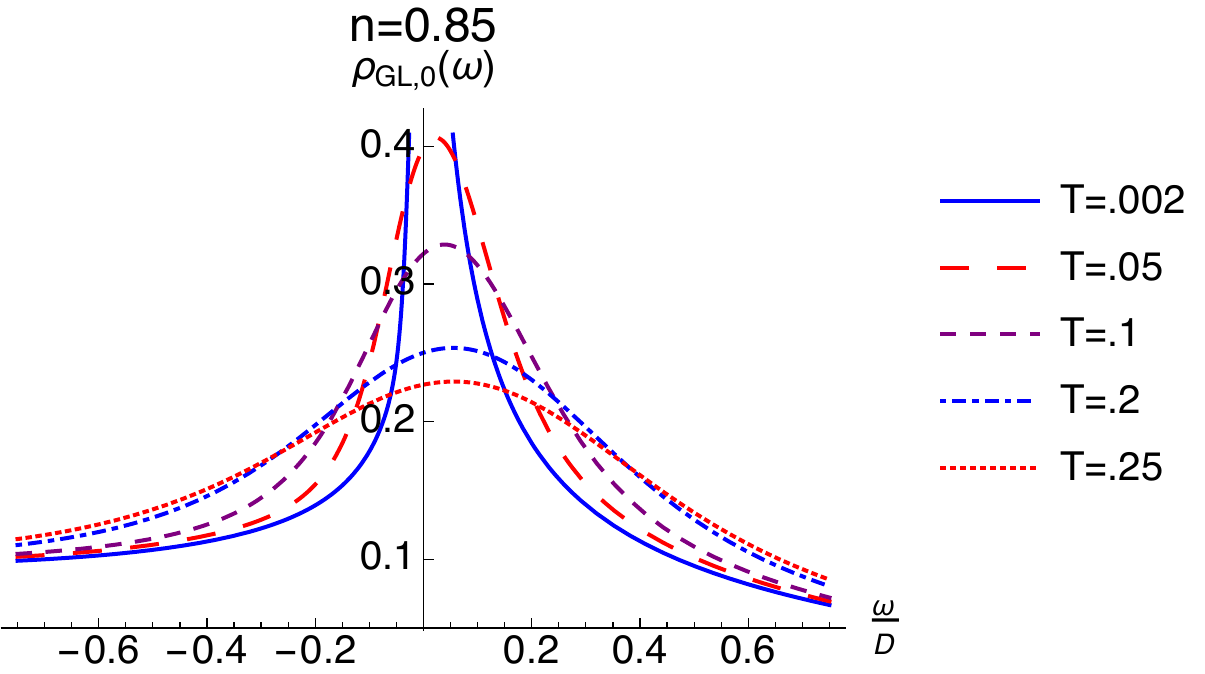}
\includegraphics[width=.95\columnwidth]{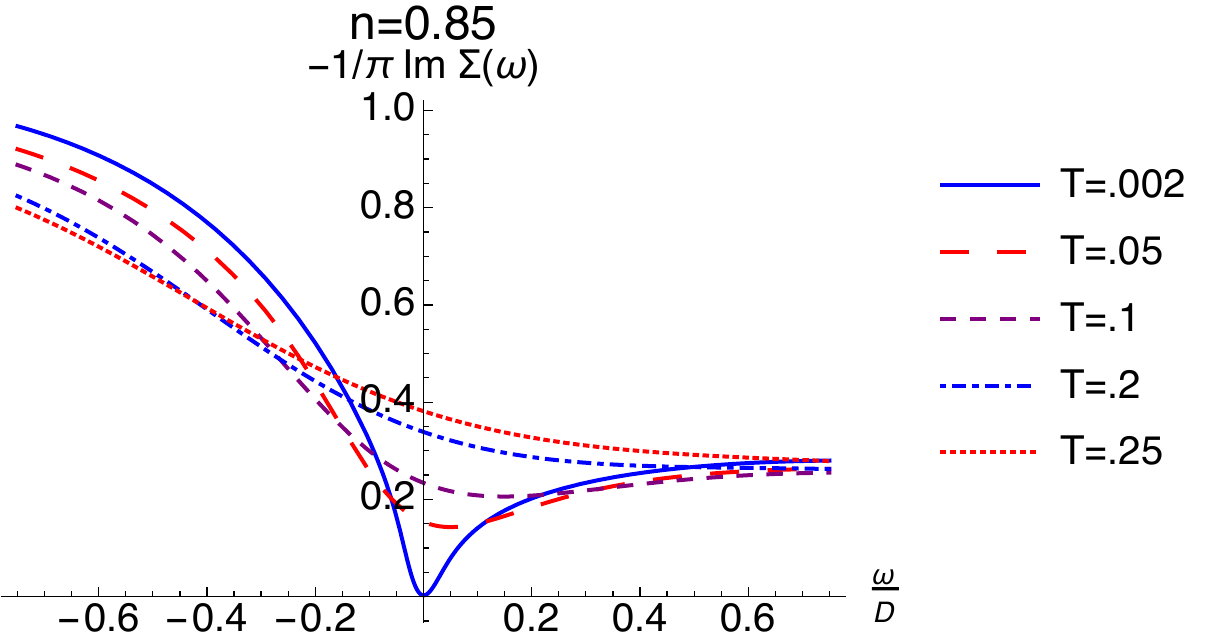}
\includegraphics[width=.95\columnwidth]{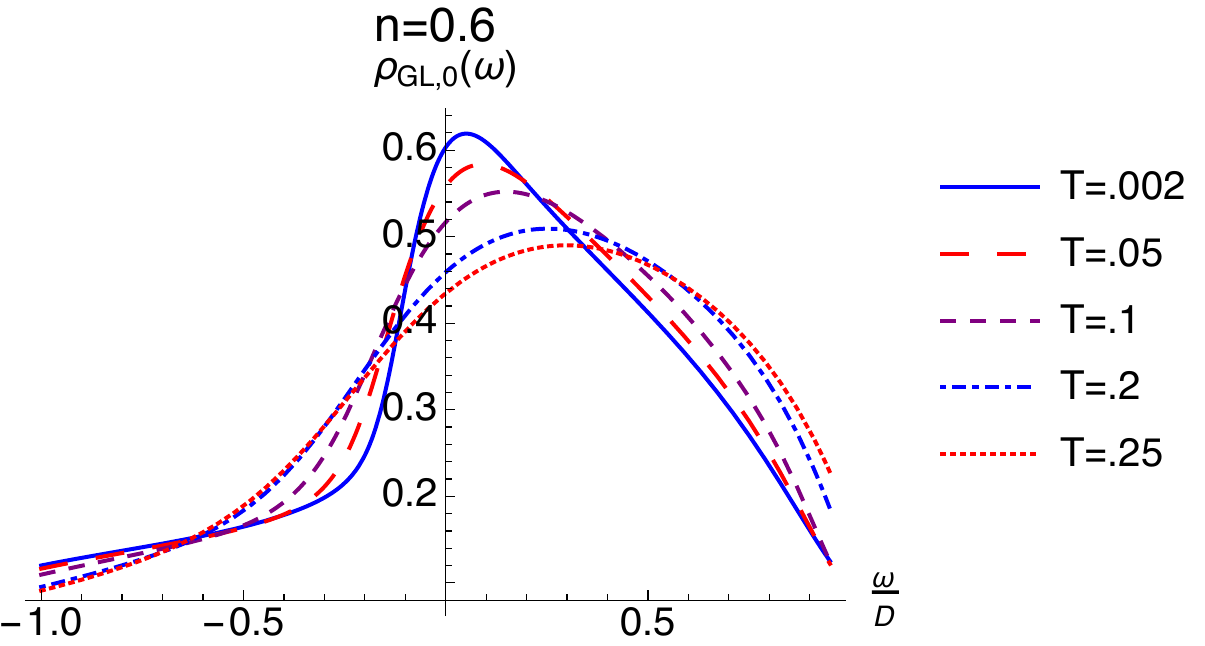}
\includegraphics[width=.95\columnwidth]{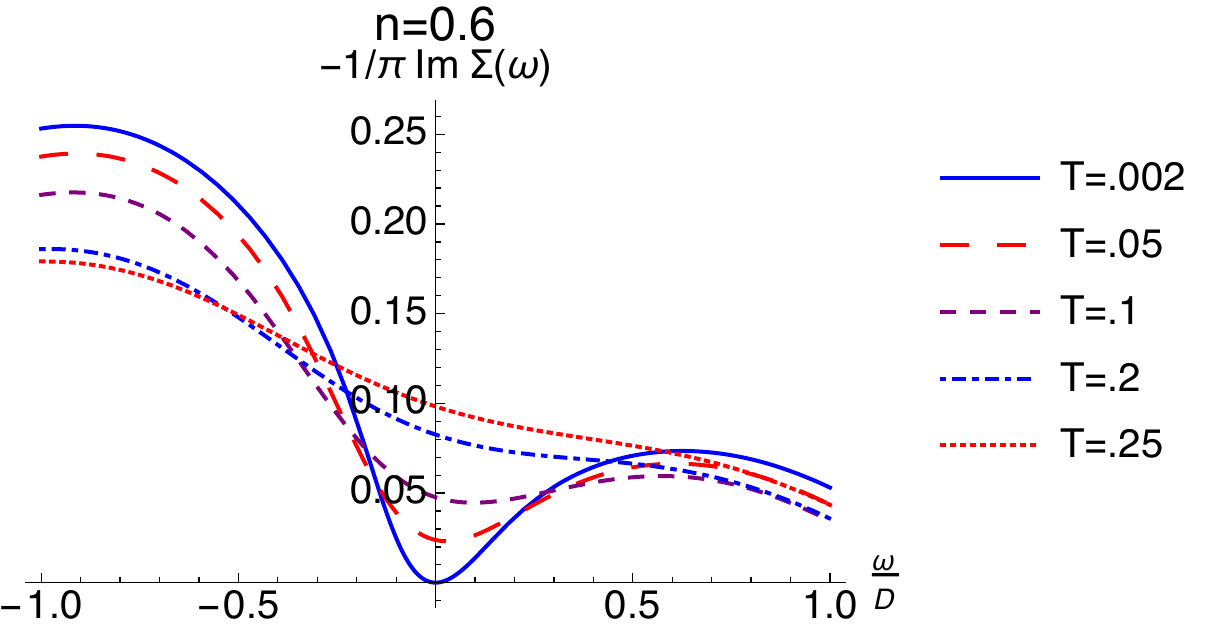}
\caption{The temperature variation  with the frequency $\omega/D$, of the  spectral function $\rho_{\G L,0}$ on the left and the Dyson self-energy  $-\frac{1}{\pi} \Im m \, \Sigma$ on the right, at density $n=.875$ (top),
 $n=0.85$ (middle) and at $n=0.6$ (bottom).  With increasing T we note the rapid broadening and shifting of  $\rho_{\G L,0}$. Here  $-\frac{1}{\pi} \Im m \, \Sigma$ displays a rapid destruction of the coherent Fermi liquid behavior observed  at the  lowest T, by the  filling up of the minimum at $\omega=0$. 
Comparing the top two sets shows that at the lowest hole-density,  a small change in T has a large effect, due to the low effective Fermi temperature.  
  We also observe here, as well as in \figdisp{LowTspectra}, that $-\frac{1}{\pi} \Im m \, \Sigma$ has a strong asymmetric correction to the quadratic $\omega$ dependence of the standard Fermi liquid, as highlighted in the bottom left panel of  \figdisp{DynamicSigma}. This is in accord with one of the basic analytical  predictions of the ECFL theory, and also is found in the DMFT results. } 
\label{HighTspectra}
\end{center}
\end{figure*}

In \figdisp{HighTspectra} we display the $T$ dependence of the spectral function and the self-energy. One of the advantages of our computational scheme is the ease with which $T$ variation can be computed. We are thus able to obtain easily the crossover from a coherent (extremely correlated) Fermi liquid regime at low $T$ to an incoherent non degenerate correlated state. The spectral function peaks rapidly broaden and shift as the temperature is increased. We also note that the Fermi coherence- signaled by a small magnitude of $\Im m \Sigma$ at small $\omega$ is rapidly lost on heating,  leading to a flat and structureless function.  A comparison of the curves at $n=0.85$ and $0.875$ show that in this range of densities, where the $Z$ is already very small, the effective Fermi temperature is also diminished since the same (small)  variation of $T$ produces a relatively large change in the damping. 

\section{Temperature-dependence of Resistivity and related quantities. \label{resist}}
Perhaps  the single  most important characterization  of a theory is via  the resistivity.  It is  a notoriously hard   object to calculate reliably, and yet one that is most sensitive to the lowest energy excitations of the system. Since we have argued that the present version of ECFL captures the low energy excitations of the electron, it is useful to examine its  results for resistivity for the \tJ model in infinite dimensions, or equivalently the $U=\infty$ Hubbard model. The resistivity 
 has been calculated numerically from DMFT quite  recently in two papers \refdisp{Badmetal,Kotliar}, and hence it is of interest to see how our analytical calculation compares with these exact results.
 
We start with the Kubo expression for resistivity, with the vertex correction thrown out, thanks to the simplification arising from $d\to \infty$:
\beq
\sigma_{DC}= \frac{2 \pi \hbar e^2}{V} \sum_k \; (v_k^x)^2\; \int d\omega \,  (-\partial f/\partial \omega)\; \rho^2_{\G}(\epsilon_k, \omega),
\eeq
where the band velocity is given as $\hbar v_k^x= \partial \varepsilon_k/\partial  k_x$. We wrap the velocity into a useful function 
\beq 
\Phi(\epsilon)&=& \frac{1}{a_0} \frac{1}{N_s} \sum_k \delta(\varepsilon-\varepsilon_k) \; (v_k^x)^2/a_0^2\nn\\
 &=& \frac{1}{a_0} D(\epsilon) \langle \frac{(v_k^x)^2}{a_0^2} \rangle_{\varepsilon_k = \epsilon}, 
\eeq
where $a_0$ is the lattice constant in the hypercubic lattice, and $N_s $ the number of sites and we use the Bethe lattice semicircular density-of-states $D(\epsilon) = \frac{2}{\pi D} \sqrt{1-\epsilon^2/D^2}$.
Deng et. al. \cite{Badmetal,footnote-IRM} calculate  that
\beq
\frac{\Phi(\epsilon)}{\Phi(0)}=  \Theta(1-\epsilon^2/D^2)\; \sqrt[3/2]{1-\epsilon^2/D^2}.
\eeq
where $\Phi(0)$ is absorbed into a constant $\sigma_0= e^2 \hbar \Phi(0)/D$, which  is identified with the Ioffe-Regel-Mott conductivity. With this choice of the vertex we obtain
\beq
 \sigma_{DC} = \sigma_{0}\times  2 \pi D \int \int\, d\epsilon  \, d \omega\,  (-\partial f/\partial \omega) \left( \frac{\Phi(\epsilon)}{\Phi(0)}\right)\;  \rho^2_{\G}(\epsilon, \omega).\nn\\
\eeq

\begin{figure*}[*htb]
\begin{center}
\includegraphics[width=.67\columnwidth]{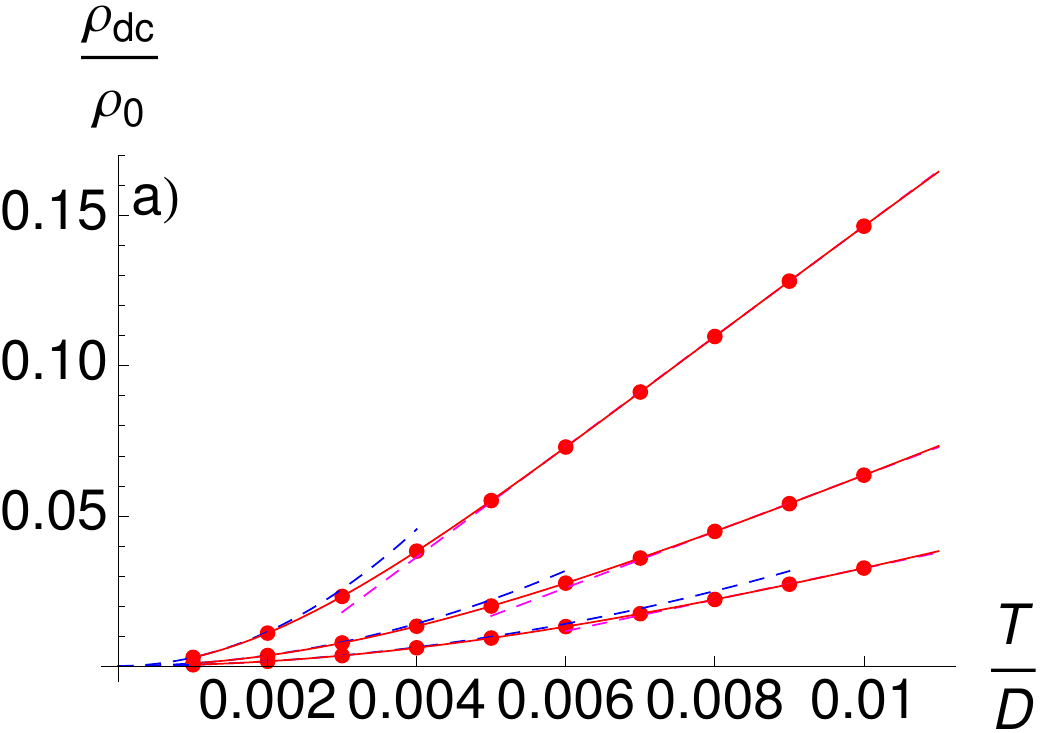}
\includegraphics[width=.67\columnwidth]{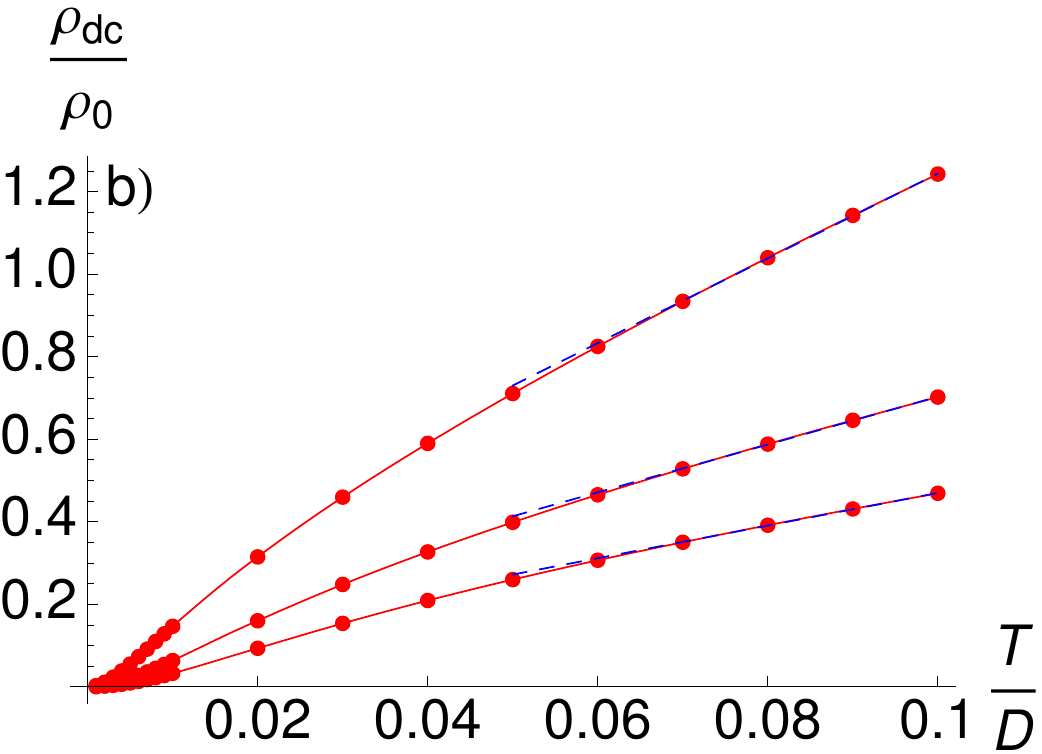}
\includegraphics[width=.7\columnwidth]{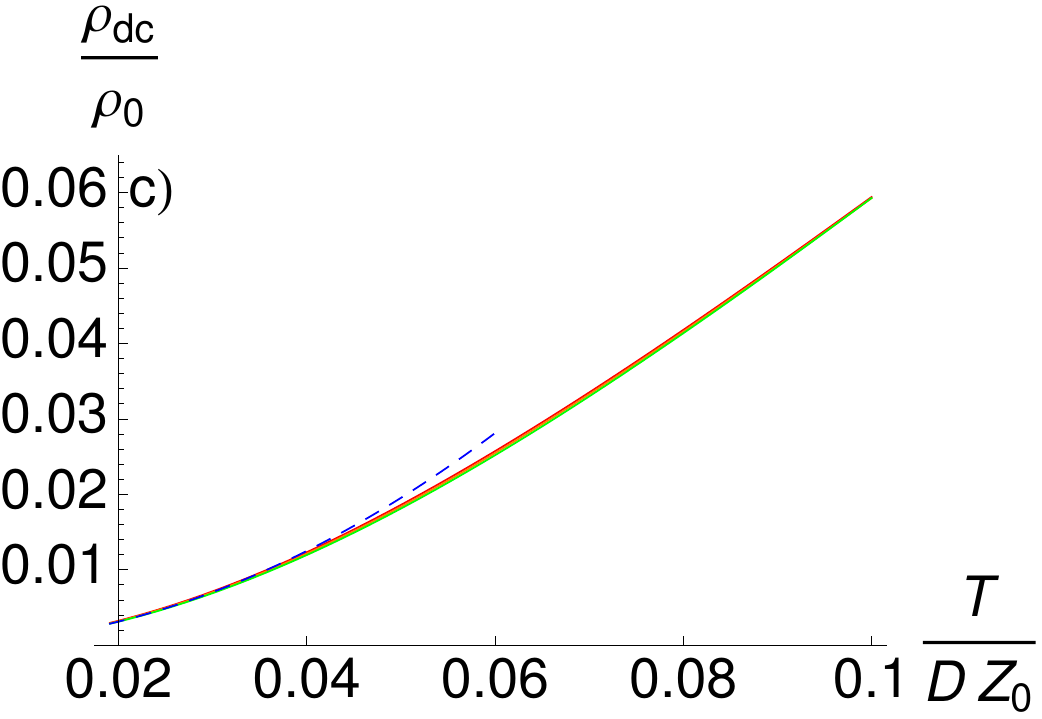}
\caption{ {\bf Panels a) and b):} $\frac{\rho_{dc}}{\rho_0}$ vs. $\frac{T}{D}$ for $\delta = .75, \ .8, \ .85$ from bottom to top. In {\bf Panel a)}, the blue dashed parabola tracks the FL regime, $0<T< T_{FL}$ where $\frac{\rho_{dc}}{\rho_0}\propto (\frac{T}{D})^2$. The magenta dashed line tracks the first linear regime, $T_{FL}<T\lesssim .01 D$. In {\bf Panel b)}, the blue dashed line tracks the second linear regime, $T\gtrsim.07D$. {\bf Panel c):} $\frac{\rho_{dc}}{\rho_0}$ vs. $\frac{T}{D Z(T=0)}$ for $\delta = .75, \ .8, \ .85$ (red, orange, green). The blue dashed parabola tracks the Fermi-liquid regime, demonstrating that $T_{FL}= (c  \ D) \times Z(T=0)$, with $c\approx.05$, and that $\frac{\rho_{dc}}{\rho_0}$ is a function of $\frac{T}{D Z(T=0)}$ for $T\lessim 2 T_{FL}$.
} 
\label{Resistivities}
\end{center}
\end{figure*}

We write the  (inverse) Greens function at real $\omega$  as
\beq
\G^{-1}_{\pm}(\epsilon,\omega) &= A(\omega) - \epsilon \pm i B(\omega),
\eeq
where the retarded  case corresponds to $\G_{+}$, and 
\beq
A(\omega,T)&=& \omega + \chem - \Re e \ \Sigma(\omega,T) \nn \\
B(\omega,T)&=& \pi \rho_{\Sigma}(\omega,T) = - \Im  m \ \Sigma(\omega,T),
\eeq
and $\Sigma$ is the Dyson self-energy. Setting $D=1$ and using the identities  $\rho_{\G}= i/(2 \pi) (\G_{-}-\G_{+})$ and  $\G^2_{\pm}=\partial_\epsilon \G_{\pm}$, and further  integrating by parts over $\epsilon$ we obtain
\beq
{\sigma}&=& \sigma_0 \times \int \ d\omega \,  (-\partial f/\partial \omega) \xi(\omega), \nn \\
\xi(\omega) &=& \frac{1}{2 \pi} \int d\epsilon  \left\{   \frac{i}{B}(\G_{+} -\G_{-})  \frac{\Phi(\epsilon)}{\Phi(0)} + (\G_{+} + \G_{-})  \frac{\Phi'(\epsilon)}{\Phi(0)}  \ \right\}.\nn\\
\label{resxi}
\eeq
Using the explicit form of $\Phi$ and $\G_{\pm}$ we re-express $\xi$  exactly as
\beq
\xi(\omega)= \frac{1}{\pi} \int_{-1}^{1} \ d\epsilon \; \sqrt{1-\epsilon^2} \times  \frac{1-3 \epsilon A + 2 \epsilon^2}{B^2+(A-\epsilon)^2}.
\eeq 
The evaluation of this integral is straightforward, and leads to a cumbersome result. A simple answer for  the leading behavior when $B\ll1$ can be found, provided $(A-\epsilon)$ goes through zero in the  interval of integration. Since we will see that $|A|\ll1$ for all temperatures and frequencies of interest ($\omega \sim 0$, $\frac{T}{D}\lessim .3$), this will always be the case. We may  write $\epsilon =A + B \tan(\theta)$, retain the leading terms for small B, and set $B\to 0$ in the remainder.   With this we obtain the 
 asymptotic   approximation:  
\beq
 \lim_{B \ll1} \xi(\omega) \sim \frac{(1-A^2(\omega))^{3/2}}{B(\omega)} \; \Theta(1-A^2(\omega)).
 \label{Resformula}
\eeq
In Fig. \ref{Resistivities}, we use \disp{Resformula} to plot $\frac{\rho_{dc}}{\rho_0}$ vs. $\frac{T}{D}$ for $.75\le\delta \le .85$, where, $\rho_0=\frac{1}{\sigma_0}$. These resistivity curves have both the same shape and the same scale as those found through DMFT. We find a Fermi-liquid regime ($\frac{\rho_{dc}}{\rho_0}\propto (\frac{T}{D})^2$) for $0<T<T_{FL}$, where $T_{FL}= (c  \ D) \times Z(T=0)$, and $c\approx.05$. Furthermore, $\frac{\rho_{dc}}{\rho_0}$ is a function of $\frac{T}{D Z(T=0)}$ for $T\lessim 2 T_{FL}$ (Fig. (\ref{Resistivities}c)). An important scale emphasized in DMFT studies \cite{Badmetal, Kotliar} is the Brinkman-Rice scale ($T_{BR}=D\delta$), which is the renormalized band-width of the quasi-particles. Since $Z(T=0)\propto \delta^\alpha$, with $\alpha>1$, the Fermi-liquid scale is contained within the Brinkman-Rice scale, and is smaller than the latter by some power of $\delta$. As $T$ is increased above $T_{FL}$, the Fermi-liquid regime is followed by a linear regime for $T_{FL}<T\lesssim .01 D$. In Fig. (\ref{Resistivities}a), the Fermi-liquid regime is tracked using the blue dashed parabola, while the linear regime is tracked using the magenta dashed line.  Finally, this linear regime connects continuously to a second linear regime, existing for $T\gtrsim.07D$ (displayed in Fig. (\ref{Resistivities}b)). 

\begin{figure*}[*htb]
\begin{center}
\includegraphics[width=.67\columnwidth]{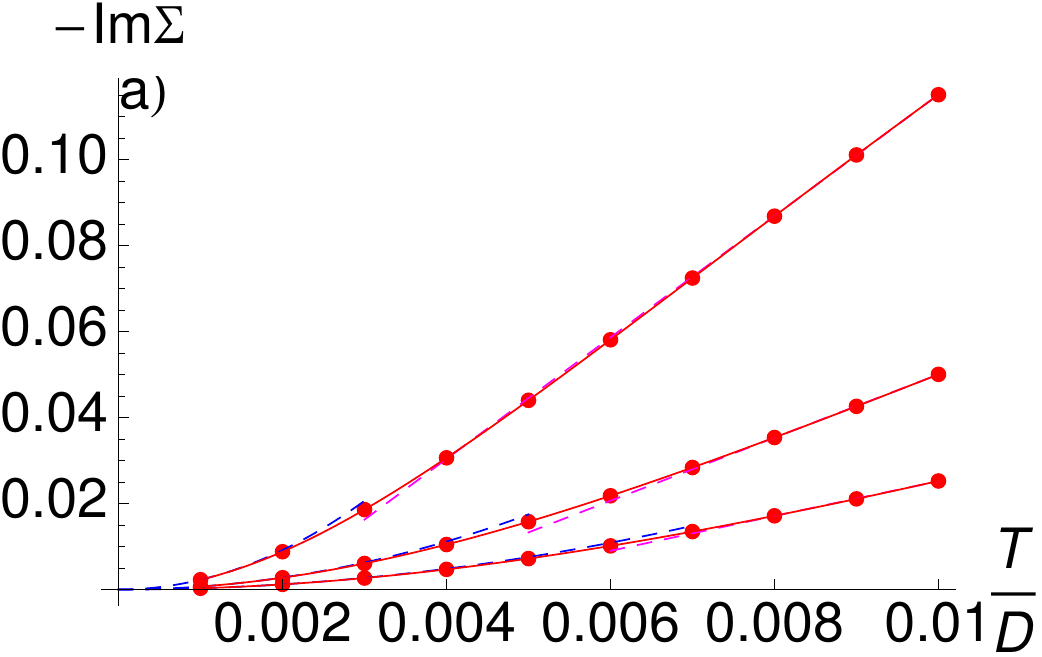}
\includegraphics[width=.67\columnwidth]{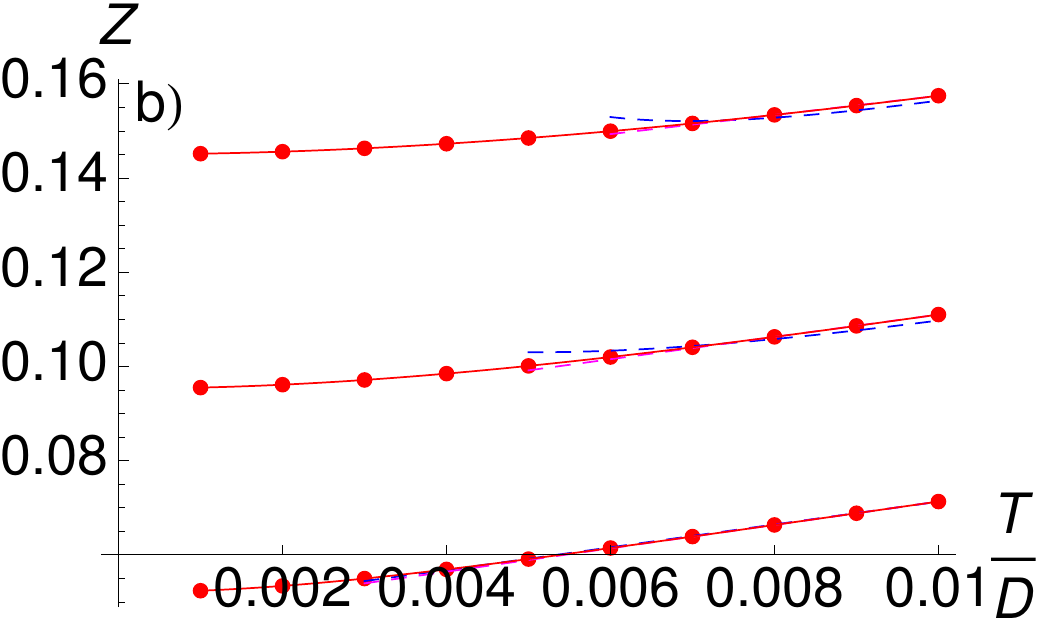}
\includegraphics[width=.7\columnwidth]{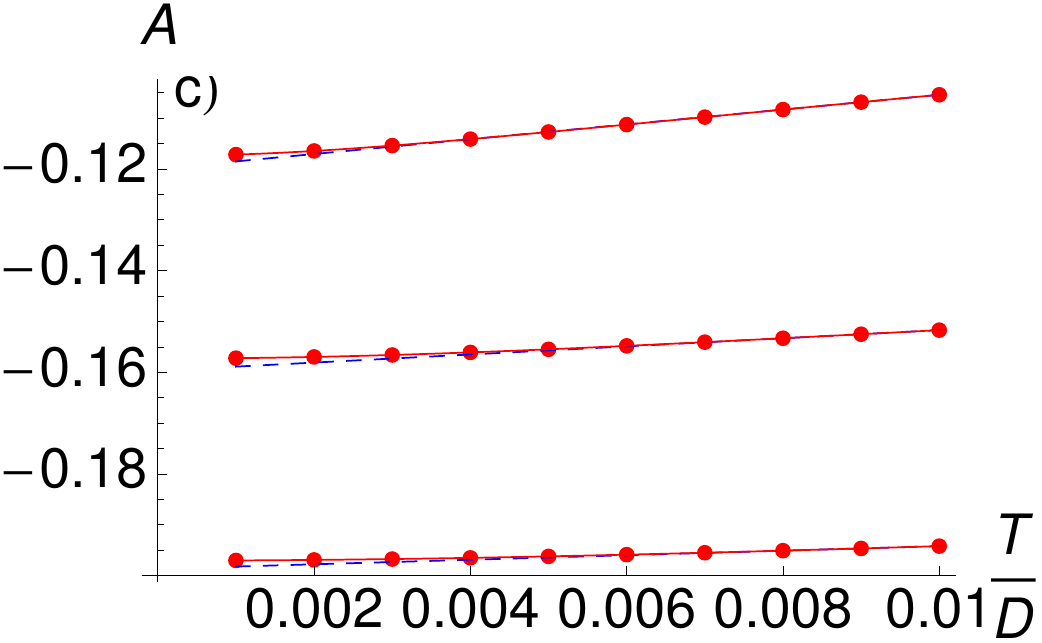}
\caption{{\bf Panel a):} ($- \Im m \,{\Sigma}(0,T)$) vs. $\frac{T}{D}$ for $\delta = .75, \ .8, \ .85$ from bottom to  top. $(-\Im m \,{\Sigma}(0,T))$ is quadratic for $T\lesssim T_{FL}$ (tracked by the blue dashed parabola) and linear for $T_{FL} \lesssim T \lesssim .01D$ (tracked by the magenta dashed line). {\bf Panel b):} $Z(T)$ vs. $\frac{T}{D}$ for $\delta = .75, \ .8, \ .85$ from top to bottom. $Z(T)$ is approximately constant for $T\lesssim T_{FL}$, and grows linearly for $T_{FL} \lesssim T \lesssim .01D$, with a slope on the order of the band-width (tracked by the magenta dashed line). The blue-dashed curve is the fit to the functional form $Z(T) = \sqrt{\frac{1 + a T + b T^2}{c + d T}}$ using a broader range of temperatures than the one shown here (Fig. (\ref{highTobjects}b)). This form works well for $T\gtrsim T_{FL}$. {\bf Panel c):} $A(0,T)=\chem(T) - \Re e \ \Sigma(0,T)$ vs. $\frac{T}{D}$ for $\delta = .75, \ .8, \ .85$
from bottom to  top. For $ T_{FL}\lesssim T\lesssim .01D$, it is linear, as tracked by the dashed blue line.
} 
\label{lowTobjects}
\end{center}
\end{figure*}

We now analyze more closely the low-temperature regime ($T\lesssim .01 D$). For this range of temperatures, the Sommerfeld expansion can be applied to \disp{resxi}. To leading order ($-\partial f/\partial \omega= \delta(\omega)$), and using \disp{Resformula}, this gives 
\beq
\rho_{DC}&\sim&\rho_0 \times \frac{ - \Im m \,{\Sigma}(0,T)}{\left( 1- \{ \chem- \Re e\,  \Sigma(0,T)\}^2\right)^{3/2}}.
\label{asymptotic}
\eeq
The constituent objects $- \Im m \,{\Sigma}(0,T)$ and $A(0,T)$ are plotted along with $Z(T)$ in the relevant temperature range in \figdisp{lowTobjects}. We first examine $A(0,T)=\chem - \Re e \ \Sigma(0,T)$, displayed in Fig. (\ref{lowTobjects}c). For $ T_{FL}\lesssim T\lesssim .01D$, it is linear, as tracked by the dashed blue line. We also notice that $A^2(0,T)\ll1$, and can therefore be neglected in \disp{asymptotic}. \disp{asymptotic} then implies that the resistivity is proportional to $(-\Im m \,{\Sigma}(0,T))$ in this low-temperature range. Accordingly, in Fig. (\ref{lowTobjects}a), we see that $(-\Im m \,{\Sigma}(0,T))$ is quadratic for $T\lesssim T_{FL}$ (tracked by the blue dashed parabola) and linear for $T_{FL} \lesssim T \lesssim .01D$ (tracked by the magenta dashed line). Finally, in Fig. (\ref{lowTobjects}b), we see that $Z(T)$ is approximately constant for $T\lesssim T_{FL}$, and grows linearly for $T_{FL} \lesssim T \lesssim .01D$, with a slope on the order of the band-width (tracked by the magenta dashed line). The blue-dashed curve tracks the functional form discussed below, which approximates $Z(T)$ very well for $T\gtrsim T_{FL}$. As emphasized in \refdisp{Kotliar}, the temperature dependence of $(-\Im m \,{\Sigma}(0,T))$ and $Z(T)$ lead to a quasi-particle scattering rate, defined as $ (-\Im m \,{\Sigma}(0,T)) \times Z(T)$, which is quadratic well above $T_{FL}$.

\begin{figure*}[*htb]
\begin{center}
\includegraphics[width=.7\columnwidth]{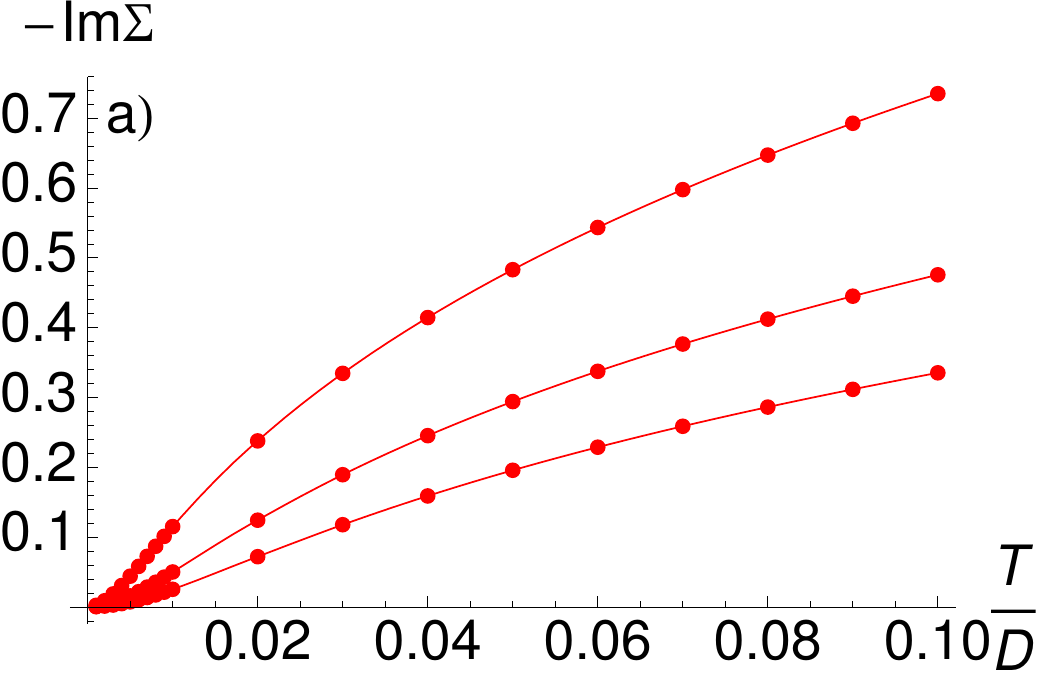}
\includegraphics[width=.67\columnwidth]{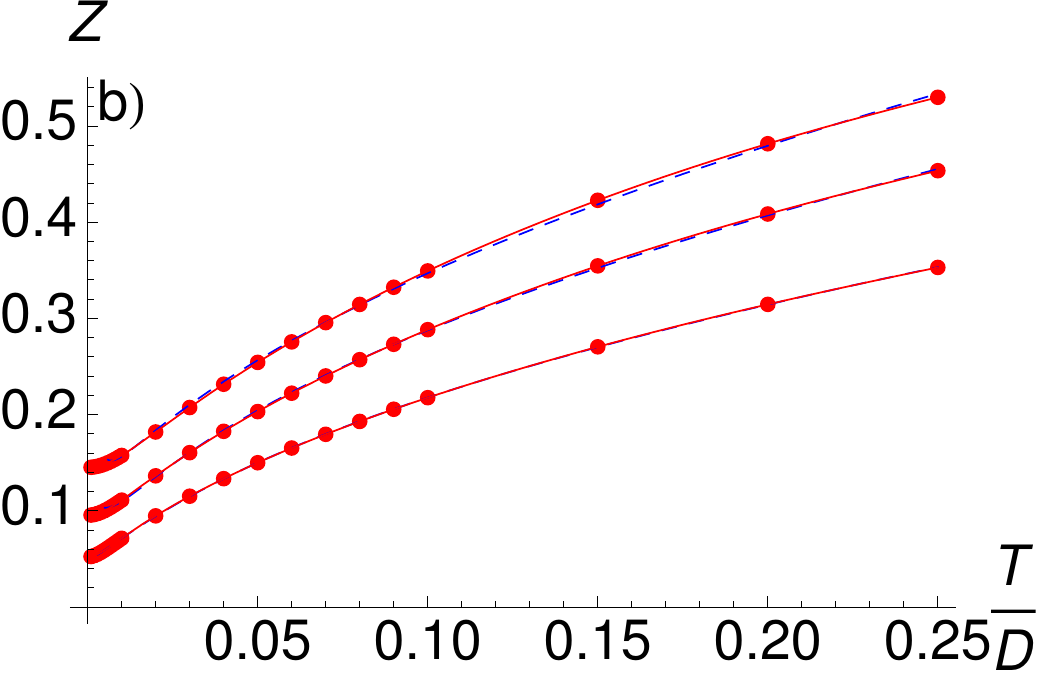}
\includegraphics[width=.67\columnwidth]{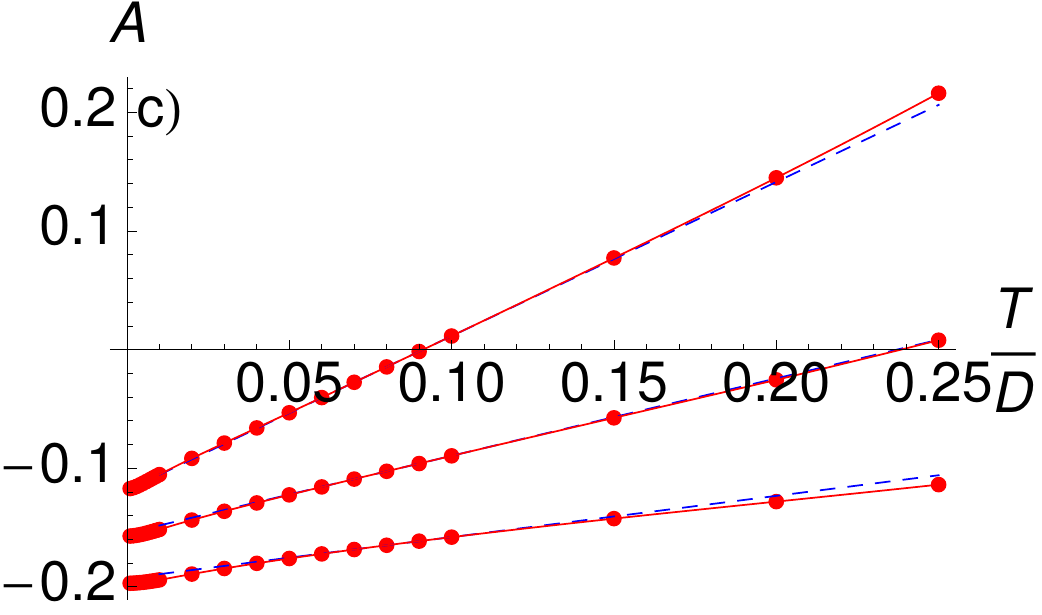}
\caption{Same plots as in \figdisp{lowTobjects} over a broader range of temperatures. {\bf Panel a):} $(-\Im m \,{\Sigma}(0,T))$ continues to grow as $T$ is increased beyond $.01D$, until it finally begins to saturate at higher temperatures. {\bf Panel b):} The blue dashed curve is the fit to the functional form $Z(T) = \sqrt{\frac{1 + a T + b T^2}{c + d T}}$, which works well for $T\gtrsim T_{FL}$. For $T\gtrsim .01 D$, $Z(T) \propto \sqrt{T}$. {\bf Panel c):} The blue dashed line tracks the second linear regime in $A(T)$ (with a slope slightly smaller than the first) for $T\gtrsim .01D$.   } 
\label{highTobjects}
\end{center}
\end{figure*}

In \figdisp{highTobjects}, we plot the temperature dependence of these objects in a broader temperature-range. In Fig. (\ref{highTobjects}c), the blue dashed line indicates the presence of a second linear regime in $A(T)$ (with a slope slightly smaller than the first), meeting the latter at a kink at $T\approx .01D$. Fig. (\ref{highTobjects}a) shows that for $T > .01D$, $(-\Im m \,{\Sigma}(0,T))$ continues to grow, until it finally begins to saturate at higher temperatures. Finally, in Fig. (\ref{highTobjects}b), we fit $Z(T)$ to the functional form $Z(T) = \sqrt{\frac{1 + a T + b T^2}{c + d T}}$, tracked by the blue dashed curve. This form works well for $T \gtrsim T_{FL}$. For $T\lesssim .01 D$, it reproduces the behavior shown in Fig. (\ref{lowTobjects}b), while for $T\gtrsim .01 D$, it is consistent with the behavior $Z(T)\propto \sqrt{T}$. Therefore, $Z^2(T)$ is linear in $T$ over a very wide temperature range starting with $T\approx .01D$.

\section{Conclusions \label{conclusions}}
{
In this work we have presented  an analytical calculation of properties of the \tJ model in infinite dimensions, and shown that it provides a quantitative description of variables known from exact numerical work in \refdisp{Badmetal} and \refdisp{Kotliar}. The results include the quasiparticle weight, the self-energies and spectral functions with particle hole asymmetry that have been argued to be characteristic of very strong correlation\cite{ECFL-Asymmetry,DMFT-ECFL}. Finally we   also give a good account of the  temperature variation of    resistivity.
Results with the present technique  at high T  are less reliable  and are not presented.  In the low to intermediate T results reported here,   we  reproduce the main features of the exact  DMFT calculations, including   a narrow regime with quadratic $T$ dependence followed by  two distinct linear T dependent  regimes. 
We are further  able to  identify  the origin of these regimes in terms of the parameters of the theory.

 The \tJ  model studied here  contains two essential ingredients of strong correlations: the physics of Gutzwiller projection to the subspace of single occupancy, and the physics of the  superexchange. The first   is captured in the present scheme, while the second  is lost,  since we   limit the study to infinite dimension for the purpose of benchmarking against known exact results. The scheme by itself has no intrinsic limitations to the  case studied, and  is  generalizable to finite dimensions as well as finite superexchange.  Thus  it  may be expected to yield interesting  results in lower dimensions, including transitions between different broken symmetry states.  Such calculations are  currently underway.

 In the present work  we have discussed the characteristics of the resulting ECFL state. The state reported here    is  Fermi liquid like, but only so at a surprisingly low temperature. Upon minimal warming,  this    state devolves into  one   exhibiting  linear resistivity. Our calculation yields a      reduction    of the   effective Fermi temperature,  due to  extreme correlations, that    far exceeds  the  expectations \cite{comment-scales-fermi} based on a simple estimate   $T^{eff}_{F}\sim \delta \,  T_{F}$.

Within  the terms of its  limitations of $d \to \infty$ and $J=0$, this work provides useful insights.  At the  density  $n\sim .85$ relevant for cuprate superconductors,  we obtain a state displaying  linear resistivity for T beyond $\sim 45$K as seen in  \figdisp{absolute-resistivity}. A similar onset of linearity occurs at a slightly higher  T    within  DMFT, the difference is due to   our $Z$ (from \figdisp{Zdelta}) being about half of the exact value. If we   imagine that the effects of reduced dimensionality and  nonzero J can stabilize this smaller onset scale, then the possibility of observing the {  asymptotic} $T^2$ resistivity of a Fermi liquid would become remote.   Thus the quadratic behavior,  so essential  for making  a {\em formal distinction}  between Fermi liquids and the elusive  non Fermi liquids\cite{Laughlin}, could be rendered unobservable in practice as well as divested of any essential difference.


\section{Acknowledgements}
We  thank Antoine Georges for useful discussions about resistivity in DMFT.
The work at UCSC was supported by the U.S. Department of Energy (DOE), Office of Science, Basic Energy Sciences (BES) under Award \# FG02-06ER46319.


\begin{thebibliography}{99}
\bibitem{eqnreference} These equations can be found in \refdisp{Edward-Sriram-2} Eqs.~(56-61) by setting the exchange $J\to0$. In particular Eq.~(65-a) gives the quoted $\Psi_{[1]}$,  Eq.~(66-a) gives the quoted $\chi_{[0]}$, and Eqs.~(66-b,67-a) give the quoted $\chi_{[1]}$. We stress that in this version the sumrule for $\gamma$ is used exactly and hence we retain only the first term  Eq.~(64-a) abandon the higher order  $\lambda$ expansion terms  
Eq.~(64-b-c-d) etc.
\bibitem{half} The other half is clearly in the occupied region $\omega <0$.
\bibitem{footnote-IRM} { The value $\Phi(0)$ can be calculated as $$\Phi(0)= \frac{D^3}{2 \pi (a_0 d \hbar^2)} \frac{\varepsilon_F}{(D^2-\varepsilon_F^2)^{3/2}}, $$ 
and yields an estimate of the IRM conductivity scale  $$\sigma_0= \left( \frac{e^2}{h } \, \frac{1}{a_0 d} \right)\times \frac{D^2  \varepsilon_F}{(D^2-\varepsilon_F^2)^{3/2} } .$$ The proportionality to inverse dimension is expected since the current flows in one of $d$ directions.  }
\bibitem{comment-scales-fermi} Using $C\equiv k_B \frac{T}{T_F}$ for a band of non-interacting electrons, where $C$ is the specific heat, we find that $T_F = \frac{3}{4\pi} \frac{D}{k_B}$. With the half bandwidth $D\sim 12000$K (1 e.V.)  a naive estimate  using $n=0.85$ and a crude formula $T_F^{eff}= \delta T_F$ would yield $T^{eff}_F\sim 430$ K. In DMFT \cite{Badmetal}, this scale is reduced to $90$ K, while in the present theory, it is reduced even further down to $30$ K.
\end{thebibliography}
\end{document}